\documentclass[aps,amsmath,amssymb,superscriptaddress,prb,10pt,twocolumn,eqrefnum]{revtex4-2}
\usepackage{amsmath}    
\usepackage{graphicx}  
\usepackage{verbatim}  
\usepackage{color}      
\usepackage{subfigure} 
\usepackage{hyperref}   
\usepackage{xcolor,txfonts,mathtools,mathrsfs}
\usepackage{lineno}
\definecolor{SC-color}{named}{red}

\definecolor{SCh-color}{named}{blue}

\newcommand{\al}{\alpha}

\newcommand{\p}{\pi}

\newcommand{\w}{\omega}

\newcommand{\De}{\Delta}
\newcommand{\G}{\Gamma}
\renewcommand{\S}{\Sigma}

\newcommand{\pd}{\partial}

\newcommand{\round}[1]{\left( #1 \right)}
\renewcommand{\square}[1]{\left[ #1 \right]}
\newcommand{\curly}[1]{\left\{#1\right\}}

\newcommand{\mat}[4]{\left(\begin{array}{cc}#1&#2\\#3&#4\end{array}\right)}

\newcommand{\beq}{\begin{equation}}
\newcommand{\eeq}{\end{equation}}
\newcommand{\Beq}{\begin{eqnarray}}
\newcommand{\Eeq}{\end{eqnarray}}
\newcommand{\bml}{\begin{multline}}

\newcommand{\bea}{\begin{align}}
\newcommand{\ena}{\end{align}}
\newcommand{\bsp}{\begin{split}}
\newcommand{\esp}{\end{split}}

\newcommand{\nn}{\nonumber}

\newcommand{\ez}{\hat{\bf z}}

\newcommand{\bk}{{\bf k}}

\newcommand{\hG}{\hat G}
\newcommand{\hg}{\hat g}

\newcommand{\cH}{{\hat{\mathcal H}}}

\newcommand{\bn}{{\bf n}}

\newcommand{\hbsig}{\hat{\boldsymbol\sigma}}

\begin{document}
\title{Controlling Orbital Magnetism of Band Electron Systems via Bath Engineering}
\author{Subrata Chakraborty}
\email[Correspondence to: ]{schrkmv@gmail.com}
\affiliation{Fachbereich Physik, Universit{\"a}t Konstanz, D-78457 Konstanz, Germany}
\affiliation{Department of Physics, Queens College of the City University of New York, Queens, New York 11367, USA}
\author{So Takei}
\email[Correspondence to:]{stakei@qc.cuny.edu}
\affiliation{Department of Physics, Queens College of the City University of New York, Queens, New York 11367, USA}
\affiliation{Physics Doctoral Program, Graduate Center of the City University of New York, New York, NY 10016, USA}
\date{\today}

\begin{abstract} 
Orbital magnetism arises due to the coherent cyclotron motion of band electrons. System-bath entanglement is expected to disrupt this coherent electronic motion and quench orbital magnetism. Here, we show that a suitably-tailored bath can lead to an enhancement of the orbital diamagnetic susceptibility of a multi-band electron system and can even convert an orbital paramagnetic response into a diamagnetic one as the system-bath coupling is increased. We also demonstrate how a van Hove singularity in the bath density of states can be exploited to generate a giant enhancement of the orbital magnetic susceptibility. Our work opens doors to the possibility of controlling the orbital magnetic response of band electron systems through bath engineering. 
\end{abstract}
\maketitle

{\it Introduction}.--The study of orbital magnetism in isolated band systems began nearly a century ago with the Landau-Peierls theory of diamagnetism in single-band systems~\cite{landauZP30,peierlsZP33}. For multi-band systems, orbital magnetism can be non-trivial due to field-mediated couplings between different Bloch bands. Wallace and McClure first predicted that such inter-band effects can lead to anomalously large diamagnetism in graphene at the charge-neutrality point~\cite{Wallace1947,McClure}. Inter-band effects were later found to give divergent paramagnetism in two-dimensional (2D) band systems when the Fermi level is tuned to a van Hove singularity in the density of states~\cite{Vignale1991}. Fukuyama first derived a useful expression for the orbital magnetic susceptibility of a multi-band system in terms of the electron Green function (GF)~\cite{Fukuyama1970,Fukuyama1971,Fukuyama2007}. His works were further developed by others~\cite{koshinoPRB07,gomezPRL11,raouxPRB15} to provide a systematic tool for computing the susceptibility of multi-band systems. Enhanced diamagnetism has been observed in Dirac materials, such as graphene and bismuth, when the chemical potential is tuned within the weak effective mass gap~\cite{raouxPRB15,Ominato2013,LiPRB2015,Suetsugu2021,vallejobustamante}. Strong orbital magnetism has also been reported in BiTeI semiconductor, where giant Rashba spin-orbit interaction mediates coupling between the valence and conduction bands~\cite{Schober2012, Suzuura2016, Bahramy2017}.

While the orbital magnetism of isolated band systems is now well-studied, the same cannot be said for electrons coupled to a dissipative bath. Recent interests in the enhanced diamagnetism of isolated materials therefore lead us to question whether the orbital magnetic response of a multi-band system can be controlled by bath engineering. Orbital magnetism is precluded within 
classical statistical mechanics by the Bohr-van Leeuwen theorem~\cite{bohrTHE11,leeuwenJPR21,vleckBOOK32}: Studies based on the quantum Langevin equation~\cite{caldeiraANP83,lindenbergBOOK90,kampenBOOK07} have indeed shown that a vanishing diamagnetic moment is obtained in the limit of very strong energy-independent dissipation~\cite{marathePRA89,liPRA90,dattaguptaPRL97,bandyopadhyayJPCM06,satpathiJSP19,alpomishevPRE21}. While dissipation, at first sight, seems to be harmful to diamagnetism, the circumstances under which it can be beneficial are counterintuitive and have not yet been identified. The goal of this work is to show how the orbital magnetic response of a multi-band system can be controlled and enhanced by opening the system to a bath with a suitably-tailored local density of states (LDOS).

In this work, we explore how the orbital magnetic susceptibility of a 2D band electron system can be controlled by entangling the system with a fermionic bath.
Using the gauge-invariant Keldysh GF formalism, we derive an expression for the susceptibility that is valid for any band structure and arbitrary bath LDOS: This work therefore generalizes the GF expression for the susceptibility of an isolated system | see, e.g., Ref.~\onlinecite{raouxPRB15} | to an open, dissipative system. Then for a generalized graphene band structure, both gapless and gapped (e.g., Sb-doped bismuth), we show how the susceptibility can be controlled using the bath. A wide-band bath with a nonzero LDOS over the bandwidth, introduces uniform decoherence and quenches the susceptibility for strong system-bath coupling. However, with engineered semiconducting and half-filled narrow bandwidth baths, which introduce non-uniform decoherence over the energy domain, we show that the susceptibility can be enhanced by increasing the system-bath coupling and can also be converted from paramagnetic to diamagnetic and vice versa. Furthermore we find that van Hove singularities in the bath LDOS can lead to giant enhancements in the susceptibility.

{\it General multi-band model}.--We begin with a general, multi-band electron Hamiltonian $H_S=\sum_{\bk}\sum_{mn}\cH^{mn}_{\bk}c^\dag_{\bk  m}c_{\bk  n}$, where $c^\dag_{\bk  m}$ ($c_{\bk  n}$) creates (annihilates) an electron with crystal momentum $\bk=(k_x,k_y)$  in band $m$ ($n$), and the band indices take the values $m,n=1,2,\dots,\mathcal{N}$~\footnote{In this work, we consider spinless electrons and focus solely on orbital magnetism.}. The system is tunnel-coupled to a fermion bath, which is assumed to be equilibrated with temperature $T$ and chemical potential $\mu$. We fix the temperature to zero; however, the main results of this work should qualitatively hold at any finite low $T$, which should merely lead to trivial thermal broadening. Ignoring feedback effects of the system on the bath, the (inverse) retarded GF for the system in the presence of the bath can be written as $[\hg_{\bk}(\w)]^{-1}=\w-\cH_{\bk}/\hbar-\hat{\S}^{R}_\bk(\w)$, where $\hat{\S}^{R}_{\bk}(\w)$ defines the retarded self-energy matrix arising due to system-bath coupling. 

Since the real part of the self-energy does not play an essential role in the following discussion, we will ignore it and write $\hat{\S}^{R}_\bk(\w)=-i\hat\G_\bk(\w)$. Within the local tunneling approximation for the system-bath coupling, the retarded self-energy matrix becomes diagonal in band-space $\hat\G_\bk(\omega)=\G(\omega)\equiv\p\rho_B(\omega)|\xi|^2/\hbar$, where $\rho_B(\omega)$ is the LDOS of the bath, and $\xi$ is the amplitude for system-bath tunneling~\cite{SM}. Here, $\Gamma(\omega)$ measures the rate at which electrons with energy $\hbar\w$ escape into the bath and introduces spectral broadening and decoherence to the band electrons~\footnote{The introduction of the dissipative bath within the local tunneling approximation does not lead to couplings between the bands; however, our work can be readily generalized to the case where the bath also induces inter-band coupling.}. 

{\it Orbital magnetic susceptibility}.--The gauge-invariant GF formalism developed by Onoda {\em et al}.~\cite{onodaPTP06} provides a general framework for systematically calculating the linear (and non-linear) response of a band electron system to a uniform, static electromagnetic field. In a uniform, static magnetic field applied perpendicular to the system, i.e., ${\bf{B}}=B\ez$, the formalism leads to the following retarded GF up to second order in $B$, \beq
\label{gRB2}
\hG_{\bk}(\w,B)=\hg_{\bk}(\w)+i\round{\frac{eB}{2\hbar}}\hat{{g}}^{(1)}_{\bk}(\w)+\round{\frac{eB}{2\hbar}}^2\hat{{g}}^{(2)}_{\bk}(\w)\,,
\eeq
where $(\w,\bk)$ is now the mechanical energy-momentum vector, and the detailed expressions for $\hat{{g}}^{(1)}_{\bk}(\w)$ and $\hat{{g}}^{(2)}_{\bk}(\w)$ are provided in the Supplemental Material~\cite{SM}. In the absence of the bath, Eq.~\eqref{gRB2} correctly reproduces the GF expression up to second order in $B$ obtained recently using a different, gauge-invariant approach~\cite{raouxPRB15}. The orbital magnetic susceptibility of the dissipative, multi-band electron system is then given by
\beq
\begin{multlined}
\chi=\frac{\mu_0e^2}{S\hbar^2}\int\frac{d\w}{2\pi}\,\Theta(\mu-\hbar \w)\\
\times\sum_{\bk}{\rm Im}\curly{\square{\hbar(\w+i\G(\w))-\mu}{\rm Tr}\square{\hat{{g}}^{(2)}_{\bk}(\w)}}\,, 
\end{multlined}
\label{june8a} 
\eeq
where $\mu_0$ is the vacuum permeability, $S$ is the area of the 2D electron system, and $\Theta(x)$ is the Heaviside step function~\cite{SM}. Equation~\eqref{june8a} may be interpreted as the generalization of the result obtained recently~\cite{raouxPRB15} for a closed, multi-band electron system to the open, dissipative case. The difference between these two cases stems from the (energy-dependent) broadening of the electron levels $\G(\w)$, which modifies the electron density of states | the Tr term in Eq.~\eqref{june8a} | as well as the energy prefactor $\hbar(\w+i\G(\w))$. The controllability of $\chi$ is rooted in this energy-dependent level broadening.  

{\it Particle-hole symmetric two-band systems}.--Let us apply Eq.~\eqref{june8a} to a particle-hole symmetric, two-band system. Its Hamiltonian can be written generally as $\hat{\mathcal{H}}_\bk={\bf{f}}_\bk\cdot\hbsig$, where $\hbsig$ is the vector of Pauli matrices in the sub-lattice space and ${\bf{f}}_\bk=(f^x_\bk,f^y_\bk,f^z_\bk)$. The energy eigenvalues are $\pm\varepsilon_\bk$, where $\varepsilon_\bk=|{\bf{f}}_\bk|$, and the energy eigenstates are characterized by the Berry curvature $\Omega_{ij}(\bk)=-\bn_\bk\cdot(\bn_\bk^i\times\bn_\bk^j)/2$, where $\bn_\bk={\bf{f}}_\bk/\varepsilon_\bk$ and $\bn_\bk^i=\pd_{k_i}\bn_\bk$, and by the contra-variant geometric tensor
\beq
\mat{M^{xx}}{M^{xy}}{M^{yx}}{M^{yy}}=\frac14\mat{\bn^y\cdot\bn^y}{-\bn^x\cdot\bn^y}{-\bn^x\cdot\bn^y}{\bn^x\cdot\bn^x}\,.
\eeq
The susceptibility Eq.~\eqref{june8a} then becomes $\chi=\chi_{\rm{LP}} +\chi_\Omega+\chi_{{M}}$, where the first term is the intra-band Landau-Peierls term~\cite{landauZP30,peierlsZP33}, and the remaining two terms give the inter-band Berry curvature and geometric tensor contributions~\cite{raouxPRB15,piechonPRB16,SM}. 

\begin{figure}[t!]
\includegraphics[width=\linewidth]{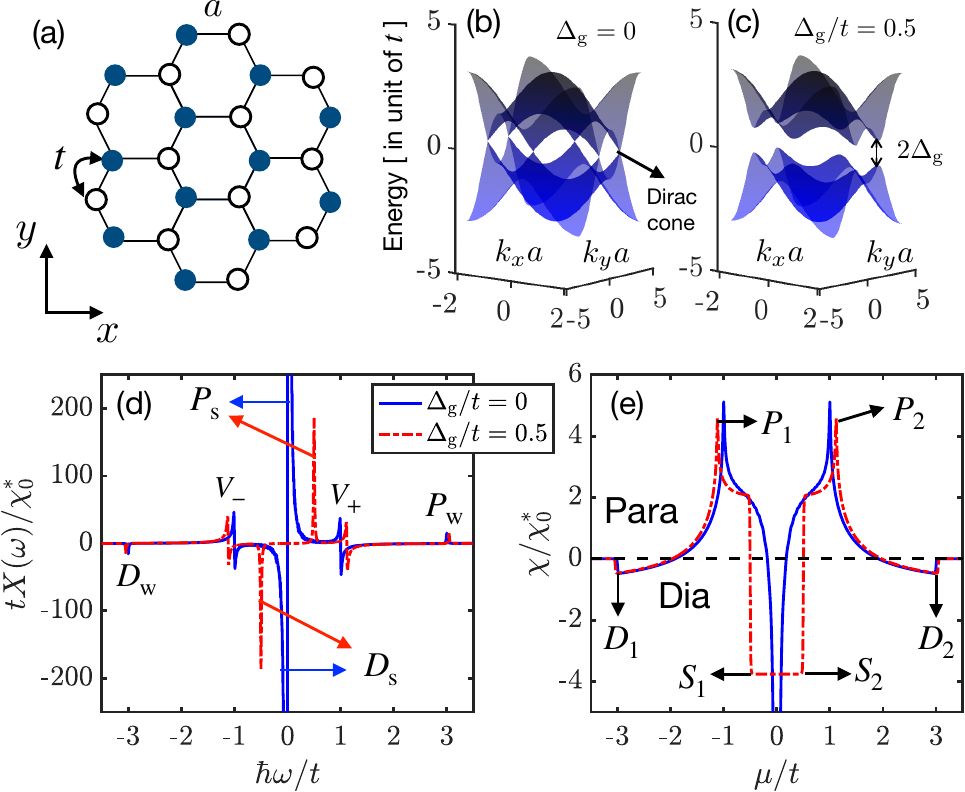}
\caption{(a) Schematic of the honeycomb lattice, where $a$ and $t$ are the lattice constant and the hopping amplitude, respectively. Panels (b) and (c) are the band dispersions for the gapless and gapped graphene systems, respectively. (d) The susceptibility spectral function $X(\w)$ for the closed system, i.e., $\G(\w)\rightarrow0^+$. (e) The corresponding orbital susceptibility $\chi(\mu)$ [see Eq.~\eqref{chimu}].
The spectral function $X(\omega)$ in (d) features a weak diamagnetic (paramagnetic) peak $D_{\rm w}$ ($P_{\rm w}$), a strong diamagnetic (paramagnetic) peak $D_{\rm{s}}$ ($P_{\rm{s}}$), and ``double-peaks" $V_\pm$ due to the van Hove singularities in the electronic density of states. The corresponding $\chi(\mu)$ in (e) features weak diamagnetic kink ($D_{1/2}$) and strong paramagnetic peak ($P_{1/2}$). For gapless graphene the $\chi(\mu)$ also features a strong orbital diamagnetic peak at $\mu=0$. Furthermore, the $\chi(\mu)$ of gapped graphene exhibits strong in-gap orbital diamagnetism between $S_1$ and $S_2$, where $-\Delta_{\rm{g}}\leq \mu \leq \Delta_{\rm{g}}$.  
}
\label{fig1}
\end{figure} 
Defining the zero-field, retarded GF in the eigenbasis (labeled by $\al=\pm 1$) as $g^{-1}_{\bk\al}(\w)=\w-\al\varepsilon_\bk/\hbar+i\G(\w)$, the components of the susceptibility read
\begin{align}
\label{LP}
\chi_{\rm LP}&=-\frac{\mu_0e^2}{12\p S\hbar^3}\int d\w\sum_{\bk\al}{\rm Im}\left[F(\w)g_\al^2\right](\varepsilon_{xx}\varepsilon_{yy}-\varepsilon_{xy}^2)\,,\\
\label{berry}
\chi_\Omega&=-\frac{\mu_0e^2}{\p S\hbar^3}\int d\w\sum_{\bk\al}{\rm Im}\left[F(\w)\left(\al\frac{\hbar g_\al}{\varepsilon}-g_\al^2\right)\right]\varepsilon^2\Omega_{xy}^2\,,\\
\label{tensor}
\chi_{{M}}&=-\frac{\mu_0e^2}{\p S\hbar^3}\int d\w\sum_{\bk\al}{\rm Im}\left[F(\w)\left(\al\frac{\hbar g_\al}{\varepsilon}-g_\al^2\right)\right]\varepsilon_{i}\varepsilon_{j} {M}^{ij}\nn\\
&\ \ +\frac{\mu_0e^2}{2\p S\hbar^3}\int d\w\sum_{\bk\al}{\rm Im}\left[F(\w) \left( \al\frac{\hbar g_\al}{\varepsilon}\right)\right]\partial_{ij} \left( \varepsilon^2{M}^{ij}  \right), 
\end{align}
where we have suppressed the $(\w,\bk)$ dependences for brevity, $\varepsilon_{i}=\pd_{k_i}\varepsilon_\bk$, and $\varepsilon_{ij}=\pd_{k_i}\pd_{k_j}\varepsilon_\bk$. The repeated indices in Eq.~\eqref{tensor} are summed over with the understanding that $\partial_{ij}=\partial_{k_i}\partial_{k_j}$. In Eqs.~\eqref{LP}-\eqref{tensor}, we have
\beq
F(\w)=\frac{\pd}{\pd\w}\left[ \frac{\Theta(\mu/\hbar -\w)~(\w-\mu/\hbar+i\G(\w))}{1+i\G^\prime(\w)} \right]\,, 
\label{F1}
\eeq
where $\G^\prime(\w)=\pd\G(\w)/\pd\w$. We note here that, in the zero dissipation limit and in the absence of inter-band coupling, we recover the Landau-Peierls formula of orbital magnetic susceptibility. 

We now apply this result to a generalized graphene band structure, where $f^{x}_\bk=t\cos(k_xa)+2t\cos(k_xa/2)\cos(\sqrt{3}k_ya/2)$, $f^{y}_\bk=-t\sin (k_xa)+2t\sin(k_xa/2)\cos(\sqrt{3}k_ya/2)$, and $f^{z}_\bk=\Delta_g$~\cite{SM}. Here, $t>0$ is the nearest-neighbor hopping amplitude on the honeycomb lattice and $a$ is the distance between nearest-neighbor atoms [see Fig.~\ref{fig1}(a)]; the band gap is $2\Delta_g$. Figures~\ref{fig1}(b) and (c) show the energy dispersions for the gapless case ($\De_g=0$) and a gapped case ($2\De_g=t$), respectively. Hereafter, $\chi$ is plotted in units of $\chi_0^*=\mu_0e^2a^2t/(12\pi\hbar^2)$.

{\it{Bath engineering}}.--To qualitatively illustrate how the susceptibility can be bath engineered, we first note that it can generally be expressed as the integral
\beq
\label{chimu}
\chi(\mu)=\hbar\int_{-\infty}^\infty d\w~\Theta(\mu-\hbar\w)X(\w)\,,
\eeq 
where we refer to $X(\w)$ as the {\em susceptibility spectral function}. A plot of the spectral function for the generalized graphene system in the absence of the bath, i.e., $\G(\w)\rightarrow0^+$, is shown in Fig.~\ref{fig1}(d), where the solid blue (dashed red) curve corresponds to the gapless $\De_g=0$ (gapped $2\De_g=t$) case~\footnote{The spectral function $X(\w)$ obeys $X(-\w)=-X(\w)$ due to particle-hole symmetry.}. The function features a weak diamagnetic (paramagnetic) peak $D_{\rm w}$ ($P_{\rm w}$), a strong diamagnetic (paramagnetic) peak $D_{\rm{s}}$ ($P_{\rm{s}}$), and ``double-peaks" $V_\pm$ due to the van Hove singularities in the electronic density of states. The corresponding susceptibilities are given by the areas under $X(\w)$ between $-\infty<\w\le\mu/\hbar$ and are plotted in Fig.~\ref{fig1}(e).
In the figure, orbital paramagnetism for certain $\mu$ appears due to magnetic field-mediated inter-band coupling~\cite{Vignale1991, raouxPRB15}.

We now open the system to the bath. If the bath LDOS is nonzero and constant over the entire bandwidth of the graphene system, the dissipation rate may be written $\G(\w)=\G_{\rm{b}}$. In this case, increasing $\G_{\rm{b}}$ broadens and washes out the sharp features in $X(\w)$ over all energies and results in the vanishing of the $\chi$ for sufficiently strong $\G_{\rm{b}}$~\cite{SM}. The latter appears to be consistent with the Bohr-van Leeuwen theorem. 

With an $\w$-dependent dissipation rate $\G(\w)$, however, decoherence can be introduced to the electrons at selected energies. Therefore, a bath with non-trivial LDOS can be used to quench the coherent cyclotron orbits at certain frequencies while leaving those at other frequencies unaffected. We may go back to Fig.~\ref{fig1}(d) and, for example, consider a bath with a robust LDOS at energies $\hbar\w\ge0$ and with zero LDOS for $\hbar\w<0$.  Such a bath is expected to diminish the susceptibility spectral weight at $\hbar\w\ge0$, leading to the quenching of the peaks at $P_{\rm s}$, $V_{+}$, and $P_{\rm w}$, and to leave the peaks at $D_{\rm s}$, $V_-$, and $D_{\rm w}$ undisturbed. With such a bath, since $\chi(\mu)$ is given by the integral of the spectral function over $-\infty<\hbar\w\leq\mu$, one may be able to suppress the strong paramagnetic response at $\mu\approx t$ and may even turn the response diamagnetic. By selectively quenching the cyclotron orbits at certain energies, one can therefore try to engineer a desired orbital magnetic response as a function of $\mu$. In what follows, we discuss the control of $\chi(\mu)$ for the generalized graphene system using two representative bath types: a semiconductor (SM) bath and a half-filled narrow bandwidth (NB) bath. We emphasize that the applicability of the idea of bath engineering is not limited to these two bath types. A new bath with a different LDOS can be used to induce a desired $\chi(\mu)$.
\begin{figure}[t!]
\includegraphics[width=1\linewidth]{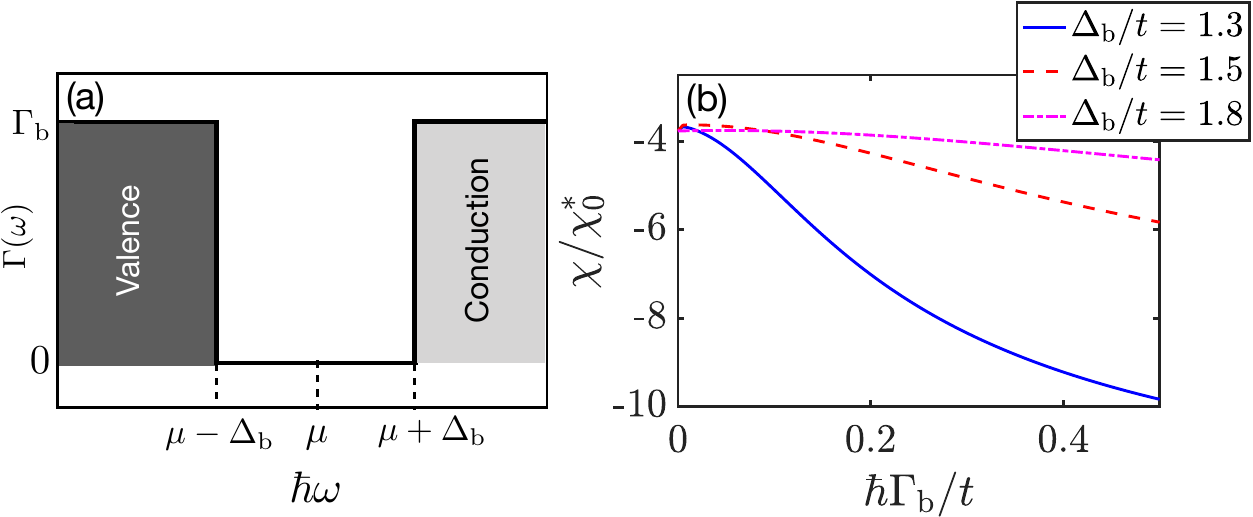}
\caption{\label{fig2}{ (a) Dissipation rate $\G(\w)$ for a semiconducting bath. The gap between the valence (dark shaded) and the conduction (light shaded) bands is $2\De_{\rm{b}}$; the chemical potential $\mu$ is set at the center of the band gap. (b) $\chi$ vs. $\G_{\rm{b}}$ for a gapped graphene system attached to the semiconducting bath: $\De_g=t/2$ and $\mu=0$ are used.}}
\end{figure}

{\em Semiconducting bath}.--The SM bath features a finite gap $2\Delta_{\rm{b}}$ between the valence and the conduction bands, and $\mu$ is located at the center of the band gap [see Fig.~\ref{fig2}(a)]. The ($\w$-dispersive) dissipation rate here can be modeled by $\G(\w)=\G_{\rm{b}}[\Theta(\mu-\Delta_{\rm{b}}-\hbar\w)+\Theta(\hbar\w -\mu-\Delta_{\rm{b}})]$, where $\G_{\rm b}$ is the strength of dissipation. We discuss how this SM bath can enhance the in-gap diamagnetism of the gapped graphene system. 

We fix the chemical potential of the bath (and of the system) to $\mu=0$. In the absence of the bath, the system at the particle-hole symmetric point is diamagnetic [see Fig.~\ref{fig1}(e)], owing to the strong diamagnetic spectral weight $D_{\rm{s}}$ [see Fig.~\ref{fig1}(d)]. However, as per Eq.~\eqref{chimu}, $\chi(\mu=0)$ is determined by the integral of $X(\w)$ over $-\infty<\hbar\w\leq0$, which consists of peaks $D_{\rm w}$, $V_-$, and $D_{\rm s}$. Though it may not be apparent from Fig.~\ref{fig1}(d), the spectral weight in the region between $D_{\rm{w}}$ and $V_-$ is slightly paramagnetic, which, when integrated over, leads to the gradual rise in paramagnetism | see the rise in the dashed red line in Fig.~\ref{fig1}(e) from $D_{\rm{1}}$ to $P_{\rm{1}}$ | and to the strong paramagnetic peak $P_{\rm{1}}$. We now place the valence band edge of the SM bath, i.e., $\mu-\De_{\rm{b}}$, somewhere in between the weak diamagnetic peak $D_{\rm{w}}$ and the van Hove point $V_-$. Such an SM bath ``filters out" the paramagnetic spectral weight between $D_{\rm{w}}$ and $\mu-\De_{\rm{b}}$ while leaving the peak $D_{\rm{s}}$ undisturbed.

Integrating the filtered spectral function, we find that the diamagnetic susceptibility is enhanced as $\G_{\rm{b}}$ is increased. This is shown in Fig.~\ref{fig2}(b), which shows that the enhancement is most pronounced when the valence band edge of the SM bath is located at $-1.3t$ because the largest amount of paramagnetic spectral weight is filtered out in this case. Similar diamagnetic enhancement can be achieved as long as the center of the SM band gap is within the band gap of the gapped graphene and the bath valence band edge is set in between $D_{\rm{w}}$ and $V_-$.

\begin{figure}[t!]
\includegraphics[width=1\linewidth]{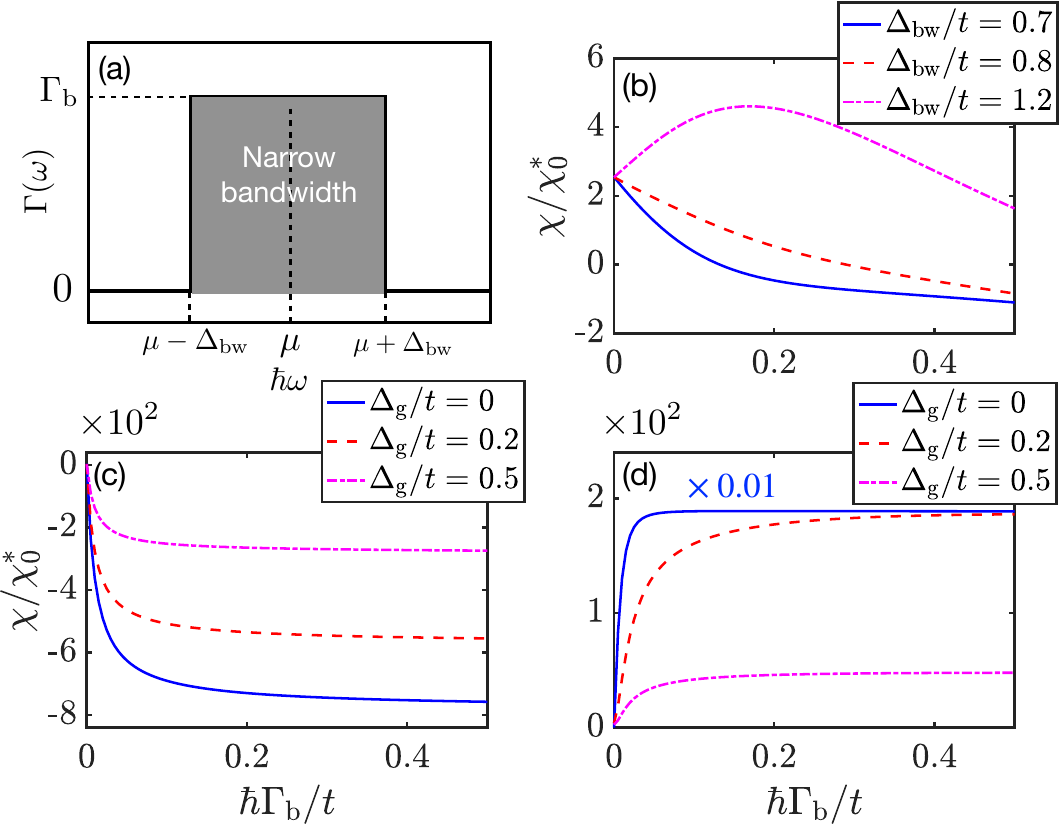}
\caption{\label{fig3}{ (a) Dissipation rate $\G(\w)$ for a half-filled narrow bandwidth bath. The bandwidth is $2\De_{\rm{bw}}$, and $\mu$ is set at the center of the band. (b)-(d) $\chi$ vs. $\G_{\rm{b}}$ for the generalized graphene attached to the half-filled narrow bandwidth bath system with $\mu=t$: (b) is obtained for the gapped case $2\De_g=t$;
(c) and (d) are obtained by aligning the lower band edge $\mu-\De_{\rm{bw}}$ with the strong diamagnetic peak $D_{\rm{s}}$ and the strong paramagnetic peak $P_{\rm{s}}$, respectively. For the gapless case $\De_g=0$, (c) and (d) are obtained by placing the edge just slightly to the left and to the right of the charge neutrality point, respectively. }}
\end{figure}

{\em Half-filled narrow bandwidth bath}.--The half-filled NB bath features a nonzero LDOS over a finite bandwidth $2\Delta_{\rm{bw}}$, and $\mu$ is located at the center of the bandwidth [see Fig.~\ref{fig3}(a)]. The dissipation rate here can be modeled by $\G(\w)=\G_{\rm{b}}[1-\Theta(\mu-\Delta_{\rm{bw}}-\hbar\w) -\Theta(\hbar\w -\mu-\Delta_{\rm{bw}})]$. We now show that a paramagnetic response of a gapped graphene system can be converted into a diamagnetic one using the half-filled NB bath. 

The chemical potential of the bath (and of the system) is set to $\mu=t$, above the system's band gap. In the absence of the bath, the system at $\mu=t$ is paramagnetic, owing to the strong paramagnetic peak $P_{\rm s}$ [see Figs.~\ref{fig1}(d) and (e)]. Once again, $\chi(\mu)$ is determined by the integral of $X(\w)$ over $-\infty<\hbar\w\leq t$, which consists of peaks $D_{\rm w}$, $V_-$, $D_{\rm s}$, and $P_{\rm s}$. If the bottom of the bath band, i.e., $\mu-\De_{\rm{bw}}$, is placed in between the strong diamagnetic peak $D_{\rm{s}}$ and the strong paramagnetic peak $P_{\rm{s}}$, the NB bath can filter out the spectral weight over the energy interval $\mu-\De_{\rm{bw}}\leq\hbar\w\leq\mu$, including the $P_{\rm{s}}$ peak, while leaving the rest of the spectrum, including the $D_{\rm{s}}$ peak, undisturbed. This drives the paramagnetic response of the system at $\mu=t$ toward a diamagnetic one as $\G_{\rm b}$ is increased.

This is illustrated by the solid blue and the dashed red curves in Fig.~\ref{fig3}(b), which shows the paramagnetic-to-diamagnetic crossover with increasing $\G_{\rm b}$ with the bottom of the bath band set at $0.3t$ and $0.2t$, respectively. Similar kind of paramagnetic-to-diamagnetic crossover can be achieved as long as $\mu-\De_{\rm{bw}}$ is set between $D_{\rm{s}}$ and $P_{\rm{s}}$ and the chemical potential satisfies $\De_g<\mu\leq t$. The general tendency for the system to turn diamagnetic is observed also for the lower band edge at $\mu-\De_{\rm{bw}}=-0.2t$, although for this case, the response is non-monotonic, where the system first becomes more paramagnetic before tending toward diamagnetism. 

{\em Van Hove Singularities}.--The non-monotonic behavior seen in Fig.~\ref{fig3}(b) for $\De_{\rm bw}/t=1.2$ cannot be explained solely by the bath-mediated spectral filtering and is rooted in the van Hove singularities of the bath LDOS. The NB bath features two van Hove singularities at $\hbar\w_\pm=\mu\pm\De_{\rm{bw}}$, where the derivatives of $\G(\w)$ are ill-defined, and these singularities play an important role in determining $\chi(\mu)$ via Eqs.~\eqref{F1} and~\eqref{chimu}. The van Hove singularity relevant for calculating $\chi(\mu)$ is at $\w_-$. When $\De_{\rm bw}/t=0.7$, $\w_-$ is situated in between $\w=0$ and the paramagnetic peak $P_{\rm s}$. The spectral weight in this region is weakly paramagnetic [see Fig.~\ref{fig1}(d)], and when $\w_-$ is in this region, it gives a strong diamagnetic contribution to $\chi$. This is seen as the initial downturn in the $\chi$ vs. $\G_{\rm b}$ curve seen for $\De_{\rm bw}/t=0.7$. For $\De_{\rm bw}/t=1.2$, $\w_-$ is located in between the diamagnetic peak $D_{\rm s}$ and $\w=0$, where the spectral weight is weakly diamagnetic. The contribution to $\chi$ coming from the van Hove point then switches sign and becomes strongly paramagnetic. This leads to the initial upturn in the $\chi$ vs. $\G_{\rm b}$ curve. All curves in Fig.~\ref{fig3}(b) tend eventually toward diamagnetism due to spectral filtering, as discussed previously. 

Interestingly, the van Hove singularities can be used to strongly enhance $\chi$. To illustrate this, we again consider the half-filled NB bath and fix $\mu=t$. Giant enhancements in the $\chi$ are shown in Figs.~\ref{fig3}(c) and (d) for three different gap sizes $\De_g=0,0.2t,0.5t$. To obtain Figs.~\ref{fig3}(c) and (d), the van Hove singularity $\w_-$ of the half-filled NB bath was aligned with the strong diamagnetic peak $D_{\rm{s}}$ and the strong paramagnetic peak $P_{\rm{s}}$, respectively, which are located at $\pm\De_g$. For $\De_g=0$, Figs.~\ref{fig3}(c) and (d) are obtained by placing $\w_-$ just slightly to the left and to the right of the charge neutrality point, respectively. The van Hove singularity is marked by diverging derivatives of $\G(\w)$. We find that when such a singularity is brought in resonance with sharp features in the susceptibility spectral function | e.g., the $D_{\rm s}$ and $P_{\rm s}$ peaks | $\chi$ can exhibit a giant enhancement. When $\w_-$ is brought in resonance with the $P_{\rm s}$ peak, we see a giant paramagnetic enhancement, while resonance with the $D_{\rm s}$ peak leads to a sign change in the susceptibility and to a giant diamagnetic enhancement. Within the current model, Figs.~\ref{fig3}(c) and (d) show that the susceptibility can be enhanced by two orders of magnitude, and in the case of $\De_g=0$ in Fig.~\ref{fig3}(d), by four orders of magnitude. These giant enhancements cannot be explained solely in terms of spectral filtering discussed previously.

{\em Conclusion}.--In this work, we theoretically analyze how the orbital magnetic susceptibility of a 2D band electron system in a perpendicular magnetic field can be controlled by a dissipative bath. A suitably-tailored bath can be used to quench the coherent cyclotron motion of the electrons at certain energies while leaving those at other energies undisturbed. Moreover, van Hove singularities in the bath density of states can also selectively amplify the orbital response of electrons whose energies are in resonance with the singularities. These bath-induced effects, together with the fact that orbital magnetism in conductors is a Fermi sea phenomenon, are at the root of how the orbital magnetic response of band electrons can be manipulated with the bath. The LDOS used in this work is idealized to illustrate the concept of bath engineering as clearly as possible; however, as a future outlook, a more realistic LDOS should be considered to quantify the susceptibility with dissipation more precisely. Also, while this work discusses how bath engineering can be used to control the orbital magnetic response of an open band electron system, it is interesting to extend this idea to control other physical properties, where electrons from the entire Fermi sea contribute to the response: This may open wider range of applications in condensed matter physics, optomechanics and in the field of open quantum systems.    

We acknowledge support by CUNY Research Foundation Project \#90922-07 10. S. T. acknowledges PSC-CUNY Research Award Program \#63515-00 51 and NSF CAREER Award DMR-2238135. S. C. acknowledges Universit{\"a}t Konstanz for providing necessary research facilities.


\begin{thebibliography}{36}%
\makeatletter
\providecommand \@ifxundefined [1]{%
 \@ifx{#1\undefined}
}%
\providecommand \@ifnum [1]{%
 \ifnum #1\expandafter \@firstoftwo
 \else \expandafter \@secondoftwo
 \fi
}%
\providecommand \@ifx [1]{%
 \ifx #1\expandafter \@firstoftwo
 \else \expandafter \@secondoftwo
 \fi
}%
\providecommand \natexlab [1]{#1}%
\providecommand \enquote  [1]{``#1''}%
\providecommand \bibnamefont  [1]{#1}%
\providecommand \bibfnamefont [1]{#1}%
\providecommand \citenamefont [1]{#1}%
\providecommand \href@noop [0]{\@secondoftwo}%
\providecommand \href [0]{\begingroup \@sanitize@url \@href}%
\providecommand \@href[1]{\@@startlink{#1}\@@href}%
\providecommand \@@href[1]{\endgroup#1\@@endlink}%
\providecommand \@sanitize@url [0]{\catcode `\\12\catcode `\$12\catcode `\&12\catcode `\#12\catcode `\^12\catcode `\_12\catcode `\%12\relax}%
\providecommand \@@startlink[1]{}%
\providecommand \@@endlink[0]{}%
\providecommand \url  [0]{\begingroup\@sanitize@url \@url }%
\providecommand \@url [1]{\endgroup\@href {#1}{\urlprefix }}%
\providecommand \urlprefix  [0]{URL }%
\providecommand \Eprint [0]{\href }%
\providecommand \doibase [0]{https://doi.org/}%
\providecommand \selectlanguage [0]{\@gobble}%
\providecommand \bibinfo  [0]{\@secondoftwo}%
\providecommand \bibfield  [0]{\@secondoftwo}%
\providecommand \translation [1]{[#1]}%
\providecommand \BibitemOpen [0]{}%
\providecommand \bibitemStop [0]{}%
\providecommand \bibitemNoStop [0]{.\EOS\space}%
\providecommand \EOS [0]{\spacefactor3000\relax}%
\providecommand \BibitemShut  [1]{\csname bibitem#1\endcsname}%
\let\auto@bib@innerbib\@empty
\bibitem [{\citenamefont {Landau}(1930)}]{landauZP30}%
  \BibitemOpen
  \bibfield  {author} {\bibinfo {author} {\bibfnamefont {L.}~\bibnamefont {Landau}},\ }\bibfield  {title} {\bibinfo {title} {Diamagnetismus der metalle},\ }\href@noop {} {\bibfield  {journal} {\bibinfo  {journal} {Zeitschrift f{\"u}r Physik}\ }\textbf {\bibinfo {volume} {64}},\ \bibinfo {pages} {629} (\bibinfo {year} {1930})}\BibitemShut {NoStop}%
\bibitem [{\citenamefont {Peierls}(1933)}]{peierlsZP33}%
  \BibitemOpen
  \bibfield  {author} {\bibinfo {author} {\bibfnamefont {R.}~\bibnamefont {Peierls}},\ }\bibfield  {title} {\bibinfo {title} {Zur theorie des diamagnetismus von leitungselektronen},\ }\href@noop {} {\bibfield  {journal} {\bibinfo  {journal} {Zeitschrift f{\"u}r Physik}\ }\textbf {\bibinfo {volume} {80}},\ \bibinfo {pages} {763} (\bibinfo {year} {1933})}\BibitemShut {NoStop}%
\bibitem [{\citenamefont {Wallace}(1947)}]{Wallace1947}%
  \BibitemOpen
  \bibfield  {author} {\bibinfo {author} {\bibfnamefont {P.~R.}\ \bibnamefont {Wallace}},\ }\bibfield  {title} {\bibinfo {title} {The band theory of graphite},\ }\href {https://doi.org/10.1103/PhysRev.71.622} {\bibfield  {journal} {\bibinfo  {journal} {Phys. Rev.}\ }\textbf {\bibinfo {volume} {71}},\ \bibinfo {pages} {622} (\bibinfo {year} {1947})}\BibitemShut {NoStop}%
\bibitem [{\citenamefont {McClure}(1956)}]{McClure}%
  \BibitemOpen
  \bibfield  {author} {\bibinfo {author} {\bibfnamefont {J.~W.}\ \bibnamefont {McClure}},\ }\bibfield  {title} {\bibinfo {title} {Diamagnetism of graphite},\ }\href {https://doi.org/10.1103/PhysRev.104.666} {\bibfield  {journal} {\bibinfo  {journal} {Phys. Rev.}\ }\textbf {\bibinfo {volume} {104}},\ \bibinfo {pages} {666} (\bibinfo {year} {1956})}\BibitemShut {NoStop}%
\bibitem [{\citenamefont {Vignale}(1991)}]{Vignale1991}%
  \BibitemOpen
  \bibfield  {author} {\bibinfo {author} {\bibfnamefont {G.}~\bibnamefont {Vignale}},\ }\bibfield  {title} {\bibinfo {title} {Orbital paramagnetism of electrons in a two-dimensional lattice},\ }\href {https://doi.org/10.1103/PhysRevLett.67.358} {\bibfield  {journal} {\bibinfo  {journal} {Phys. Rev. Lett.}\ }\textbf {\bibinfo {volume} {67}},\ \bibinfo {pages} {358} (\bibinfo {year} {1991})}\BibitemShut {NoStop}%
\bibitem [{\citenamefont {Fukuyama}\ and\ \citenamefont {Kubo}(1970)}]{Fukuyama1970}%
  \BibitemOpen
  \bibfield  {author} {\bibinfo {author} {\bibfnamefont {H.}~\bibnamefont {Fukuyama}}\ and\ \bibinfo {author} {\bibfnamefont {R.}~\bibnamefont {Kubo}},\ }\bibfield  {title} {\bibinfo {title} {Interband effects on magnetic susceptibility. ii. diamagnetism of bismuth},\ }\href {https://doi.org/10.1143/JPSJ.28.570} {\bibfield  {journal} {\bibinfo  {journal} {J. Phys. Soc. Jpn}\ }\textbf {\bibinfo {volume} {28}},\ \bibinfo {pages} {570} (\bibinfo {year} {1970})}\BibitemShut {NoStop}%
\bibitem [{\citenamefont {Fukuyama}(1971)}]{Fukuyama1971}%
  \BibitemOpen
  \bibfield  {author} {\bibinfo {author} {\bibfnamefont {H.}~\bibnamefont {Fukuyama}},\ }\bibfield  {title} {\bibinfo {title} {{Theory of Orbital Magnetism of Bloch Electrons: Coulomb Interactions}},\ }\href {https://doi.org/10.1143/PTP.45.704} {\bibfield  {journal} {\bibinfo  {journal} {Prog. Theor. Phys.}\ }\textbf {\bibinfo {volume} {45}},\ \bibinfo {pages} {704} (\bibinfo {year} {1971})}\BibitemShut {NoStop}%
\bibitem [{\citenamefont {Fukuyama}(2007)}]{Fukuyama2007}%
  \BibitemOpen
  \bibfield  {author} {\bibinfo {author} {\bibfnamefont {H.}~\bibnamefont {Fukuyama}},\ }\bibfield  {title} {\bibinfo {title} {Anomalous orbital magnetism and hall effect of massless fermions in two dimension},\ }\href {https://doi.org/10.1143/JPSJ.76.043711} {\bibfield  {journal} {\bibinfo  {journal} {Journal of the Physical Society of Japan}\ }\textbf {\bibinfo {volume} {76}},\ \bibinfo {pages} {043711} (\bibinfo {year} {2007})},\ \Eprint {https://arxiv.org/abs/https://doi.org/10.1143/JPSJ.76.043711} {https://doi.org/10.1143/JPSJ.76.043711} \BibitemShut {NoStop}%
\bibitem [{\citenamefont {Koshino}\ and\ \citenamefont {Ando}(2007)}]{koshinoPRB07}%
  \BibitemOpen
  \bibfield  {author} {\bibinfo {author} {\bibfnamefont {M.}~\bibnamefont {Koshino}}\ and\ \bibinfo {author} {\bibfnamefont {T.}~\bibnamefont {Ando}},\ }\bibfield  {title} {\bibinfo {title} {Orbital diamagnetism in multilayer graphenes: Systematic study with the effective mass approximation},\ }\href {https://doi.org/10.1103/PhysRevB.76.085425} {\bibfield  {journal} {\bibinfo  {journal} {Phys. Rev. B}\ }\textbf {\bibinfo {volume} {76}},\ \bibinfo {pages} {085425} (\bibinfo {year} {2007})}\BibitemShut {NoStop}%
\bibitem [{\citenamefont {G\'omez-Santos}\ and\ \citenamefont {Stauber}(2011)}]{gomezPRL11}%
  \BibitemOpen
  \bibfield  {author} {\bibinfo {author} {\bibfnamefont {G.}~\bibnamefont {G\'omez-Santos}}\ and\ \bibinfo {author} {\bibfnamefont {T.}~\bibnamefont {Stauber}},\ }\bibfield  {title} {\bibinfo {title} {Measurable lattice effects on the charge and magnetic response in graphene},\ }\href {https://doi.org/10.1103/PhysRevLett.106.045504} {\bibfield  {journal} {\bibinfo  {journal} {Phys. Rev. Lett.}\ }\textbf {\bibinfo {volume} {106}},\ \bibinfo {pages} {045504} (\bibinfo {year} {2011})}\BibitemShut {NoStop}%
\bibitem [{\citenamefont {Raoux}\ \emph {et~al.}(2015)\citenamefont {Raoux}, \citenamefont {Pi\'echon}, \citenamefont {Fuchs},\ and\ \citenamefont {Montambaux}}]{raouxPRB15}%
  \BibitemOpen
  \bibfield  {author} {\bibinfo {author} {\bibfnamefont {A.}~\bibnamefont {Raoux}}, \bibinfo {author} {\bibfnamefont {F.}~\bibnamefont {Pi\'echon}}, \bibinfo {author} {\bibfnamefont {J.-N.}\ \bibnamefont {Fuchs}},\ and\ \bibinfo {author} {\bibfnamefont {G.}~\bibnamefont {Montambaux}},\ }\bibfield  {title} {\bibinfo {title} {Orbital magnetism in coupled-bands models},\ }\href {https://doi.org/10.1103/PhysRevB.91.085120} {\bibfield  {journal} {\bibinfo  {journal} {Phys. Rev. B}\ }\textbf {\bibinfo {volume} {91}},\ \bibinfo {pages} {085120} (\bibinfo {year} {2015})}\BibitemShut {NoStop}%
\bibitem [{\citenamefont {Ominato}\ and\ \citenamefont {Koshino}(2013)}]{Ominato2013}%
  \BibitemOpen
  \bibfield  {author} {\bibinfo {author} {\bibfnamefont {Y.}~\bibnamefont {Ominato}}\ and\ \bibinfo {author} {\bibfnamefont {M.}~\bibnamefont {Koshino}},\ }\bibfield  {title} {\bibinfo {title} {Orbital magnetism of graphene flakes},\ }\href {https://doi.org/10.1103/PhysRevB.87.115433} {\bibfield  {journal} {\bibinfo  {journal} {Phys. Rev. B}\ }\textbf {\bibinfo {volume} {87}},\ \bibinfo {pages} {115433} (\bibinfo {year} {2013})}\BibitemShut {NoStop}%
\bibitem [{\citenamefont {Li}\ \emph {et~al.}(2015)\citenamefont {Li}, \citenamefont {Chen}, \citenamefont {Meng}, \citenamefont {Guo}, \citenamefont {Huang}, \citenamefont {Liu}, \citenamefont {Wang},\ and\ \citenamefont {Chen}}]{LiPRB2015}%
  \BibitemOpen
  \bibfield  {author} {\bibinfo {author} {\bibfnamefont {Z.}~\bibnamefont {Li}}, \bibinfo {author} {\bibfnamefont {L.}~\bibnamefont {Chen}}, \bibinfo {author} {\bibfnamefont {S.}~\bibnamefont {Meng}}, \bibinfo {author} {\bibfnamefont {L.}~\bibnamefont {Guo}}, \bibinfo {author} {\bibfnamefont {J.}~\bibnamefont {Huang}}, \bibinfo {author} {\bibfnamefont {Y.}~\bibnamefont {Liu}}, \bibinfo {author} {\bibfnamefont {W.}~\bibnamefont {Wang}},\ and\ \bibinfo {author} {\bibfnamefont {X.}~\bibnamefont {Chen}},\ }\bibfield  {title} {\bibinfo {title} {Field and temperature dependence of intrinsic diamagnetism in graphene: Theory and experiment},\ }\href {https://doi.org/10.1103/PhysRevB.91.094429} {\bibfield  {journal} {\bibinfo  {journal} {Phys. Rev. B}\ }\textbf {\bibinfo {volume} {91}},\ \bibinfo {pages} {094429} (\bibinfo {year} {2015})}\BibitemShut {NoStop}%
\bibitem [{\citenamefont {Suetsugu}\ \emph {et~al.}(2021)\citenamefont {Suetsugu}, \citenamefont {Kitagawa}, \citenamefont {Kariyado}, \citenamefont {Rost}, \citenamefont {Nuss}, \citenamefont {M\"uhle}, \citenamefont {Ogata},\ and\ \citenamefont {Takagi}}]{Suetsugu2021}%
  \BibitemOpen
  \bibfield  {author} {\bibinfo {author} {\bibfnamefont {S.}~\bibnamefont {Suetsugu}}, \bibinfo {author} {\bibfnamefont {K.}~\bibnamefont {Kitagawa}}, \bibinfo {author} {\bibfnamefont {T.}~\bibnamefont {Kariyado}}, \bibinfo {author} {\bibfnamefont {A.~W.}\ \bibnamefont {Rost}}, \bibinfo {author} {\bibfnamefont {J.}~\bibnamefont {Nuss}}, \bibinfo {author} {\bibfnamefont {C.}~\bibnamefont {M\"uhle}}, \bibinfo {author} {\bibfnamefont {M.}~\bibnamefont {Ogata}},\ and\ \bibinfo {author} {\bibfnamefont {H.}~\bibnamefont {Takagi}},\ }\bibfield  {title} {\bibinfo {title} {Giant orbital diamagnetism of three-dimensional dirac electrons in ${\mathrm{sr}}_{3}\mathrm{PbO}$ antiperovskite},\ }\href {https://doi.org/10.1103/PhysRevB.103.115117} {\bibfield  {journal} {\bibinfo  {journal} {Phys. Rev. B}\ }\textbf {\bibinfo {volume} {103}},\ \bibinfo {pages} {115117} (\bibinfo {year} {2021})}\BibitemShut {NoStop}%
\bibitem [{\citenamefont {Vallejo~Bustamante}(2023)}]{vallejobustamante}%
  \BibitemOpen
  \bibfield  {author} {\bibinfo {author} {\bibfnamefont {J.}~\bibnamefont {Vallejo~Bustamante}},\ }\emph {\bibinfo {title} {{Singular orbital diamagnetism and paramagnetism in graphene}}},\ \href {https://theses.hal.science/tel-04032004} {\bibinfo {type} {Theses}},\ \bibinfo  {school} {{Universit{\'e} Paris-Saclay}} (\bibinfo {year} {2023})\BibitemShut {NoStop}%
\bibitem [{\citenamefont {Schober}\ \emph {et~al.}(2012)\citenamefont {Schober}, \citenamefont {Murakawa}, \citenamefont {Bahramy}, \citenamefont {Arita}, \citenamefont {Kaneko}, \citenamefont {Tokura},\ and\ \citenamefont {Nagaosa}}]{Schober2012}%
  \BibitemOpen
  \bibfield  {author} {\bibinfo {author} {\bibfnamefont {G.~A.~H.}\ \bibnamefont {Schober}}, \bibinfo {author} {\bibfnamefont {H.}~\bibnamefont {Murakawa}}, \bibinfo {author} {\bibfnamefont {M.~S.}\ \bibnamefont {Bahramy}}, \bibinfo {author} {\bibfnamefont {R.}~\bibnamefont {Arita}}, \bibinfo {author} {\bibfnamefont {Y.}~\bibnamefont {Kaneko}}, \bibinfo {author} {\bibfnamefont {Y.}~\bibnamefont {Tokura}},\ and\ \bibinfo {author} {\bibfnamefont {N.}~\bibnamefont {Nagaosa}},\ }\bibfield  {title} {\bibinfo {title} {Mechanisms of enhanced orbital dia- and paramagnetism: Application to the rashba semiconductor bitei},\ }\href {https://doi.org/10.1103/PhysRevLett.108.247208} {\bibfield  {journal} {\bibinfo  {journal} {Phys. Rev. Lett.}\ }\textbf {\bibinfo {volume} {108}},\ \bibinfo {pages} {247208} (\bibinfo {year} {2012})}\BibitemShut {NoStop}%
\bibitem [{\citenamefont {Suzuura}\ and\ \citenamefont {Ando}(2016)}]{Suzuura2016}%
  \BibitemOpen
  \bibfield  {author} {\bibinfo {author} {\bibfnamefont {H.}~\bibnamefont {Suzuura}}\ and\ \bibinfo {author} {\bibfnamefont {T.}~\bibnamefont {Ando}},\ }\bibfield  {title} {\bibinfo {title} {Theory of magnetic response in two-dimensional giant rashba system},\ }\href {https://doi.org/10.1103/PhysRevB.94.085303} {\bibfield  {journal} {\bibinfo  {journal} {Phys. Rev. B}\ }\textbf {\bibinfo {volume} {94}},\ \bibinfo {pages} {085303} (\bibinfo {year} {2016})}\BibitemShut {NoStop}%
\bibitem [{\citenamefont {Bahramy}\ and\ \citenamefont {Ogawa}(2017)}]{Bahramy2017}%
  \BibitemOpen
  \bibfield  {author} {\bibinfo {author} {\bibfnamefont {M.~S.}\ \bibnamefont {Bahramy}}\ and\ \bibinfo {author} {\bibfnamefont {N.}~\bibnamefont {Ogawa}},\ }\bibfield  {title} {\bibinfo {title} {Bulk rashba semiconductors and related quantum phenomena},\ }\href {https://doi.org/https://doi.org/10.1002/adma.201605911} {\bibfield  {journal} {\bibinfo  {journal} {Advanced Materials}\ }\textbf {\bibinfo {volume} {29}},\ \bibinfo {pages} {1605911} (\bibinfo {year} {2017})}\BibitemShut {NoStop}%
\bibitem [{\citenamefont {Bohr}(1911)}]{bohrTHE11}%
  \BibitemOpen
  \bibfield  {author} {\bibinfo {author} {\bibfnamefont {N.}~\bibnamefont {Bohr}},\ }\emph {\bibinfo {title} {Studier over Metallernes Elektrontheori}},\ \href@noop {} {Ph.D. thesis},\ \bibinfo  {school} {University of Copenhagen} (\bibinfo {year} {1911})\BibitemShut {NoStop}%
\bibitem [{\citenamefont {{van Leeuwen}}(1921)}]{leeuwenJPR21}%
  \BibitemOpen
  \bibfield  {author} {\bibinfo {author} {\bibfnamefont {H.~J.}\ \bibnamefont {{van Leeuwen}}},\ }\bibfield  {title} {\bibinfo {title} {Probl\`emes de la th\'eorie \'electronique du magn\'etisme},\ }\href@noop {} {\bibfield  {journal} {\bibinfo  {journal} {J. Phys. Radium}\ }\textbf {\bibinfo {volume} {2}},\ \bibinfo {pages} {361} (\bibinfo {year} {1921})}\BibitemShut {NoStop}%
\bibitem [{\citenamefont {{Van Vleck}}(1932)}]{vleckBOOK32}%
  \BibitemOpen
  \bibfield  {author} {\bibinfo {author} {\bibfnamefont {J.~H.}\ \bibnamefont {{Van Vleck}}},\ }\href {https://books.google.com/books?id=C5o2rgEACAAJ} {\emph {\bibinfo {title} {The Theory of Electric and Magnetic Susceptibilities}}}\ (\bibinfo  {publisher} {Oxford University Press},\ \bibinfo {address} {London},\ \bibinfo {year} {1932})\BibitemShut {NoStop}%
\bibitem [{\citenamefont {Caldeira}\ and\ \citenamefont {Leggett}(1983)}]{caldeiraANP83}%
  \BibitemOpen
  \bibfield  {author} {\bibinfo {author} {\bibfnamefont {A.}~\bibnamefont {Caldeira}}\ and\ \bibinfo {author} {\bibfnamefont {A.}~\bibnamefont {Leggett}},\ }\bibfield  {title} {\bibinfo {title} {Quantum tunnelling in a dissipative system},\ }\href {https://doi.org/http://dx.doi.org/10.1016/0003-4916(83)90202-6} {\bibfield  {journal} {\bibinfo  {journal} {Ann. Phys.}\ }\textbf {\bibinfo {volume} {149}},\ \bibinfo {pages} {374 } (\bibinfo {year} {1983})}\BibitemShut {NoStop}%
\bibitem [{\citenamefont {Lindenberg}\ and\ \citenamefont {West}(1990)}]{lindenbergBOOK90}%
  \BibitemOpen
  \bibfield  {author} {\bibinfo {author} {\bibfnamefont {K.}~\bibnamefont {Lindenberg}}\ and\ \bibinfo {author} {\bibfnamefont {B.~J.}\ \bibnamefont {West}},\ }\href@noop {} {\emph {\bibinfo {title} {The Nonequilibrium Statistical Mechanics of Open and Closed}}}\ (\bibinfo  {publisher} {VCH},\ \bibinfo {address} {New York},\ \bibinfo {year} {1990})\BibitemShut {NoStop}%
\bibitem [{\citenamefont {{van Kampen}}(2007)}]{kampenBOOK07}%
  \BibitemOpen
  \bibfield  {author} {\bibinfo {author} {\bibfnamefont {N.~G.}\ \bibnamefont {{van Kampen}}},\ }\href@noop {} {\emph {\bibinfo {title} {Stochastic Processes in Physics and Chemistry}}}\ (\bibinfo  {publisher} {Elsevier Science},\ \bibinfo {address} {Amsterdam},\ \bibinfo {year} {2007})\BibitemShut {NoStop}%
\bibitem [{\citenamefont {Marathe}(1989)}]{marathePRA89}%
  \BibitemOpen
  \bibfield  {author} {\bibinfo {author} {\bibfnamefont {Y.}~\bibnamefont {Marathe}},\ }\bibfield  {title} {\bibinfo {title} {Dissipative quantum dynamics of a charged particle in a magnetic field},\ }\href@noop {} {\bibfield  {journal} {\bibinfo  {journal} {Phys. Rev. A}\ }\textbf {\bibinfo {volume} {39}},\ \bibinfo {pages} {5927} (\bibinfo {year} {1989})}\BibitemShut {NoStop}%
\bibitem [{\citenamefont {Li}\ \emph {et~al.}(1990)\citenamefont {Li}, \citenamefont {Ford},\ and\ \citenamefont {O'Connell}}]{liPRA90}%
  \BibitemOpen
  \bibfield  {author} {\bibinfo {author} {\bibfnamefont {X.~L.}\ \bibnamefont {Li}}, \bibinfo {author} {\bibfnamefont {G.~W.}\ \bibnamefont {Ford}},\ and\ \bibinfo {author} {\bibfnamefont {R.~F.}\ \bibnamefont {O'Connell}},\ }\bibfield  {title} {\bibinfo {title} {Magnetic-field effects on the motion of a charged particle in a heat bath},\ }\href@noop {} {\bibfield  {journal} {\bibinfo  {journal} {Phys. Rev. A}\ }\textbf {\bibinfo {volume} {41}},\ \bibinfo {pages} {5287} (\bibinfo {year} {1990})}\BibitemShut {NoStop}%
\bibitem [{\citenamefont {Dattagupta}\ and\ \citenamefont {Singh}(1997)}]{dattaguptaPRL97}%
  \BibitemOpen
  \bibfield  {author} {\bibinfo {author} {\bibfnamefont {S.}~\bibnamefont {Dattagupta}}\ and\ \bibinfo {author} {\bibfnamefont {J.}~\bibnamefont {Singh}},\ }\bibfield  {title} {\bibinfo {title} {Landau diamagnetism in a dissipative and confined system},\ }\href@noop {} {\bibfield  {journal} {\bibinfo  {journal} {Phys. Rev. Lett.}\ }\textbf {\bibinfo {volume} {79}},\ \bibinfo {pages} {961} (\bibinfo {year} {1997})}\BibitemShut {NoStop}%
\bibitem [{\citenamefont {Bandyopadhyay}\ and\ \citenamefont {Dattagupta}(2006)}]{bandyopadhyayJPCM06}%
  \BibitemOpen
  \bibfield  {author} {\bibinfo {author} {\bibfnamefont {M.}~\bibnamefont {Bandyopadhyay}}\ and\ \bibinfo {author} {\bibfnamefont {S.}~\bibnamefont {Dattagupta}},\ }\bibfield  {title} {\bibinfo {title} {Landau--drude diamagnetism: fluctuation, dissipation and decoherence},\ }\href@noop {} {\bibfield  {journal} {\bibinfo  {journal} {J. Phys.: Condens. Matter}\ }\textbf {\bibinfo {volume} {18}},\ \bibinfo {pages} {10029} (\bibinfo {year} {2006})}\BibitemShut {NoStop}%
\bibitem [{\citenamefont {Satpathi}\ and\ \citenamefont {Sinha}(2019)}]{satpathiJSP19}%
  \BibitemOpen
  \bibfield  {author} {\bibinfo {author} {\bibfnamefont {U.}~\bibnamefont {Satpathi}}\ and\ \bibinfo {author} {\bibfnamefont {S.}~\bibnamefont {Sinha}},\ }\bibfield  {title} {\bibinfo {title} {Non-equilibrium quantum langevin dynamics of orbital diamagnetic moment},\ }\href@noop {} {\bibfield  {journal} {\bibinfo  {journal} {J. Stat. Mech.}\ }\textbf {\bibinfo {volume} {2019}},\ \bibinfo {pages} {063106} (\bibinfo {year} {2019})}\BibitemShut {NoStop}%
\bibitem [{\citenamefont {Alpomishev}\ \emph {et~al.}(2021)\citenamefont {Alpomishev}, \citenamefont {Adamian},\ and\ \citenamefont {Antonenko}}]{alpomishevPRE21}%
  \BibitemOpen
  \bibfield  {author} {\bibinfo {author} {\bibfnamefont {E.~K.}\ \bibnamefont {Alpomishev}}, \bibinfo {author} {\bibfnamefont {G.~G.}\ \bibnamefont {Adamian}},\ and\ \bibinfo {author} {\bibfnamefont {N.~V.}\ \bibnamefont {Antonenko}},\ }\bibfield  {title} {\bibinfo {title} {Orbital diamagnetism of two-dimensional quantum systems in a dissipative environment: Non-markovian effect and application to graphene},\ }\href@noop {} {\bibfield  {journal} {\bibinfo  {journal} {Phys. Rev. E}\ }\textbf {\bibinfo {volume} {104}},\ \bibinfo {pages} {054120} (\bibinfo {year} {2021})}\BibitemShut {NoStop}%
\bibitem [{Note1()}]{Note1}%
  \BibitemOpen
  \bibinfo {note} {In this work, we consider spinless electrons and focus solely on orbital magnetism.}\BibitemShut {Stop}%
\bibitem [{SM()}]{SM}%
  \BibitemOpen
  \href@noop {} {}\bibinfo {organization} {See Supplemental Material.}\BibitemShut {Stop}%
\bibitem [{Note2()}]{Note2}%
  \BibitemOpen
  \bibinfo {note} {The introduction of the dissipative bath within the local tunneling approximation does not lead to couplings between the bands; however, our work can be readily generalized to the case where the bath also induces inter-band coupling.}\BibitemShut {Stop}%
\bibitem [{\citenamefont {Onoda}\ \emph {et~al.}(2006)\citenamefont {Onoda}, \citenamefont {Sugimoto},\ and\ \citenamefont {Nagaosa}}]{onodaPTP06}%
  \BibitemOpen
  \bibfield  {author} {\bibinfo {author} {\bibfnamefont {S.}~\bibnamefont {Onoda}}, \bibinfo {author} {\bibfnamefont {N.}~\bibnamefont {Sugimoto}},\ and\ \bibinfo {author} {\bibfnamefont {N.}~\bibnamefont {Nagaosa}},\ }\bibfield  {title} {\bibinfo {title} {Theory of non-equilibirum states driven by constant electromagnetic fields},\ }\href@noop {} {\bibfield  {journal} {\bibinfo  {journal} {Prog. Theor. Phys.}\ }\textbf {\bibinfo {volume} {116}},\ \bibinfo {pages} {61} (\bibinfo {year} {2006})}\BibitemShut {NoStop}%
\bibitem [{\citenamefont {Pi\'echon}\ \emph {et~al.}(2016)\citenamefont {Pi\'echon}, \citenamefont {Raoux}, \citenamefont {Fuchs},\ and\ \citenamefont {Montambaux}}]{piechonPRB16}%
  \BibitemOpen
  \bibfield  {author} {\bibinfo {author} {\bibfnamefont {F.}~\bibnamefont {Pi\'echon}}, \bibinfo {author} {\bibfnamefont {A.}~\bibnamefont {Raoux}}, \bibinfo {author} {\bibfnamefont {J.-N.}\ \bibnamefont {Fuchs}},\ and\ \bibinfo {author} {\bibfnamefont {G.}~\bibnamefont {Montambaux}},\ }\bibfield  {title} {\bibinfo {title} {Geometric orbital susceptibility: Quantum metric without berry curvature},\ }\href@noop {} {\bibfield  {journal} {\bibinfo  {journal} {Phys. Rev. B}\ }\textbf {\bibinfo {volume} {94}},\ \bibinfo {pages} {134423} (\bibinfo {year} {2016})}\BibitemShut {NoStop}%
\bibitem [{Note3()}]{Note3}%
  \BibitemOpen
  \bibinfo {note} {The spectral function $X(\omega )$ obeys $X(-\omega )=-X(\omega )$ due to particle-hole symmetry.}\BibitemShut {Stop}%
\end{thebibliography}
\end{document}


\title{Supplemental Material for ``Controlling Orbital Magnetism of Band Electron Systems via Bath Engineering"}
\author{Subrata Chakraborty}
\email[Correspondence to: ]{schrkmv@gmail.com}
\affiliation{Fachbereich Physik, Universit{\"a}t Konstanz, D-78457 Konstanz, Germany}
\affiliation{Department of Physics, Queens College of the City University of New York, Queens, NY 11367, USA}
\author{So Takei}
\email[Correspondence to:]{stakei@qc.cuny.edu}
\affiliation{Department of Physics, Queens College of the City University of New York, Queens, NY 11367, USA}
\affiliation{Physics Doctoral Program, Graduate Center of the City University of New York, New York, NY 10016, USA}
\date{\today}

\begin{abstract} 

\end{abstract}
\maketitle


\section{Self-energy due to system-bath coupling}
\label{decoherence}
Let us begin by studying how the local system-bath tunneling causes decoherence of a two-dimensional (2D) electron gas by its dissipative coupling to a fermionic bath. The central system is modeled as a general Gaussian theory of electrons, whose a single-particle state is labeled by canonical momentum $\bk\equiv (k_x,k_y)$ (which corresponds to the physical momentum if the system has continuous translational symmetry and to the crystal momentum for a system with lattice translational symmetry), and any remaining quantum numbers $m(n)=1,2,3....\mathcal{N}$ (could be band indices or sub-lattice indices). The system Hamiltonian may then be written as $H_S=\sum_{\bk}\sum_{mn} {\hat{\mathcal{H}}}^{mn}_{\bk}c^\dag_{\bk  m }c_{\bk  n }$. 

We model the bath as a ``perfect," fermionic bath, that is, if an electron from the system leaks into the bath, its information is immediately and irretrievably lost and it never re-enters the system after a coherent propagation. This also means that all tunneling processes at the system-bath interface are completely uncorrelated. One way to model this situation is to make infinite identical copies of the bath, one for each site of the system, such that each site of the electron system is coupled to its own independent bath. All of the baths are assumed to be identical and possess a common temperature $T$ and chemical potential $\mu$. 
Then the sites of the system we can label by the index $\bi$, such that $c_{\bi  m}=N^{-1/2} \sum_{\bk} c_{\bk  m } ~e^{i\bk\cdot \bR_{\bi}}$ where $\bR_\bi$ is the position vector of the site $\bi$ and $N$ is the total number of sites with remaining quantum number $m$ (could be band index or sub-lattice index). The bath and system-bath interaction Hamiltonians can respectively be defined by
\begin{align}
H_B&=\sum_{\bi m}\sum_{\bq}E_{\bq}d^\dag_{\bq,\bi m}d_{\bq,\bi m}~~~~{\rm{and}}~~~~ H_{SB}=\sum_{\bi m}\sum_{\bq}\round{\xi d^\dag_{\bq,\bi m}c_{\bi m}+ \xi^*c^\dag_{\bi m}d_{\bq,\bi m}} \label{bath}
\end{align}
where $\xi$ parametrizes the strength of the local system-bath tunneling. Once the bath degrees of freedom are integrated out, this introduces a self-energy term to the electron Green function of the system. The self-energy matrix in the Keldysh space can be given by the causality structure 
\beq
\check{\S}_{i m;jn}(\w) = \de_{mn}\de_{ij}|\xi|^2\sum_\bq\mat{\fg^R_{\bq;B}(\w)}{2\fg^<_{\bq;B}(\w)}{0}{\fg^A_{\bq;B}(\w)}\,,
\eeq
where $i(j)$ is the site index of the system and $m(n)$ defines the band or sub-lattices indices.
The bath electron Green functions are
\beq
\fg^{R,A}_{\bq;B}=\left[{\w-E_{\bq}/\hbar\pm i0^+}\right]^{-1}\,,\ \ \ \fg^<_{\bp}= \left(\fg^{A}_{\bq;B} -\fg^{R}_{\bq;B} \right) n_F(\w)\,.
\eeq
where $n_F(\w)=(e^{(\hbar\w-\mu)/k_BT}+1)^{-1}$ is the Fermi-Dirac distribution, $T$ is the temperature of the bath (which is equal to the temperature of the electron system when the two subsystems are in thermal equilibrium), $\mu$ is the chemical potential of the bath (which is equal to the chemical potential of the electron system when the two subsystems are in chemical equilibrium). At steady state system' s temperature and chemical potential would be equal to bath's temperature $T$ and bath's chemical potential $\mu$, respectively.
In evaluating the self-energy, we ignore the real parts of the retarded and advanced Green functions. We then obtain the self-energy matrix element in the Keldysh space as
\beq
\check{\S}_{j m;j'n}(\w) = \de_{mn}\de_{jj'} \mat{-i\G(\w)}{4i \G(\w) n_F(\w)}{0}{i \G(\w)}
\equiv \mat{{\S}^R_{j m;j'n}(\w)}{2 {\S}^<_{j m;j'n}(\w) }{0}{{\S}^A_{j m;j'n}(\w) } \label{oct27a} \,,
\eeq
where $\G(\w)=\pi(|\xi|^2/\hbar)\rho_B(\w)$ and $rho_B(\w)=-({\p\hbar})^{-1}\sum_\bq{\rm Im}\{\fg^R_{\bq;B}(\w)\}$
is the density of states of the reservoir for spin $\s$ electrons. Here, $\G(\w)$
can be interpreted as the rate at which electron in state $\ket{i\s m}$ tunnels out of the system into the bath. Under system-bath local tunneling approximation here we notice the self-energy matrix is diagonal in site-index, spin and band (or sub-lattice) space, and diagonal elements do not depend on site index $i$ and band (or sub-lattice) index $m$. 

Due to the local system-bath tunneling approximation the dressed retarded GF of the system can be expressed as
\Beq
&& \hat{g}^{R}_{\bk} =\left[{\w \hat{I}_{\mathcal{N}} -{\hat{\mathcal{H}}}_{\bk}/\hbar-\hat\S^R_{\bk}(\w)}\right]^{-1} \label{oct26a} \\
\rm{with,} && \hat\S^R_{\bk}=-i\hat\G_{\bk}(\w) = -i\G(\w) \hat{I}_{\mathcal{N}}, \label{oct26b}
\Eeq
where last equality in Eq.~\eqref{oct26b} is due to the consideration of all identical baths and identical system-bath interactions. 
$\hat{I}_{\mathcal{N}}$ is $\mathcal{N}\times\mathcal{N}$ dimensional identity matrix with $\mathcal{N}$ being the total number of bands or sub-lattices available to the system. For brevity we may suppress $\hat{I}_{\mathcal{N}}$ from the dressed GF expression of the system and denote 
\beq
 \hat{g}^{R}_{\bk} =\left[{\w  -{\hat{\mathcal{H}}}_{\bk}/\hbar +i\G(\w)}\right]^{-1}. \label{oct26c} 
 \eeq


\section{Gauge-Invariant Electron Green functions}
In what follows we use the gauge-invariant Green function formalism developed by Onoda {\it{et. al.}}~\cite{onodaPTP06}.
To study orbital magnetism, we use the gauge-invariant formulation of the quantum Boltzmann equation (QBE) to solve for the electron Green function in presence of external static and homogeneous electric and magnetic fields. In this formulation, we introduce the standard mechanical momentum $p^\mu \rightarrow k^\mu +\frac{e}{\hbar} A^\mu(\br,t)\equiv (\nu, \bp)$
with $\bp \equiv (p_x,p_y)$
where $e>0$ is the electron charge, $k^\mu\equiv (\w,\bk)$ is the canonical energy-momentum vector and $A^\mu(\br,t)\equiv (\phi(\br,t),\bA(\br,t))$ is the usual scalar-vector potentials describing external static and homogeneous electromagnetic fields. We adopt the following causality structure for the system electron Green functions and self-energies,
\beq
\cG_{\bp}(\n)=\mat{\hG^R_{\bp}(\n)}{2\hG^<_{\bp}(\n)}{0}{\hG^A_{\bp}(\n)}\ ,\ \ \ \cSig_{\bp}(\n)=\mat{\hSig^R_{\bp}(\n)}{2\hSig^<_{\bp}(\n)}{0}{\hSig^A_{\bp}(\n)}\ ,
\eeq
where the ``check" denotes matrices in the Keldysh-internal quantum number space, and the ``hat" denotes matrices only in the internal quantum number space. To account for the dissipative bath, we follow Eq.~\eqref{oct27a} and introduce the following self-energy,
\beq
\cSig_{\bp}(\n) \equiv \cSig(\n) =\mat{-i\hat\G(\n)}{4i \hat\G(\n) n_F(\n)}{0}{i \hat\G(\n)}\,,
\eeq
where $\mu$ is the chemical potential of the bath (which is equal to the chemical potential of the electron system when the two subsystems are in chemical equilibrium), and $\hat\G(\n)$ defines system-bath dissipation matrix which has dimensionality $\mathcal{N}\times \mathcal{N}$ with $\mathcal{N}$ being the total number of bands or types of lattice sites available to the system. Local system-bath tunneling approximation by Eq.~\eqref{bath} would result diagonal dissipation matrix $\hat\G(\n)=\G(\n)\hat{I}_{\mathcal{N}}$ with $\hat{I}_{\mathcal{N}}$ being the $\mathcal{N}\times \mathcal{N}$ dimensional identity matrix. We assume that the rate of dissipation is independent of spin $\s$. 

In the presence of an electromagnetic field, the left and the right Dyson equations for each spin projection read
\begin{align}
\label{ldeqem}
\square{\left(\n-\frac{\hat{\mathcal{H}}_{\bp}}{\hbar} \right)\otimes \hat{I}_K -\cSig(\n)}\star\cG_{\bp}(\n)&=1~~~{\rm{and}}~~~ \cG_{\bp}(\n)\star\square{\left(\n-\frac{\hat{\mathcal{H}}_{\bp}}{\hbar}\right)\otimes \hat{I}_K-\cSig(\n)}=1,
\end{align}
where $\hat{I}_K$ is the identity matrix in the Keldysh space and the $\star$ is the Moyal product
$[A\star B]_\bp(\n)=A_\bp(\n)e^{\frac{ie}{2\hbar}F^{\mu\nu}\overleftarrow{\pd}_{p^\mu}\overrightarrow{\pd}_{p^\nu}}B_\bp(\n)$
and
$F^{\mu\nu}=\pd_{x_\mu}A^\nu-\pd_{x_\nu}A^\mu$, where $\pd_{x_\mu}=(-\pd_t,\pd_\br)$ and
$ \pd_{\br}=(\pd_x,\pd_y)$.
In this convention for static in-plane electric ($\bE=E_x\hat{\bx} +E_y\hat{\by}$) and out of plane magnetic ($\bB=B\hat{\bf{z}}$) fields, the $2+1$-dimensional electromagnetic tensor is given by
\beq
F^{\mu\nu}=\matt{0}{E_x}{E_y}{-E_x}{0}{B}{-E_y}{-B}{0}\ .
\eeq
In our present work $E_x=E_y=0$.

\subsection{First-order correction}
Following Onoda {\it{et. al.}}~\cite{onodaPTP06} we first compute the first-order correction of the Green function due to $B$. Since we are in equilibrium, it suffices to compute only the retarded Green function. The first-order correction of the retarded Green function is
\begin{align}
\hG^{R(1)}&=-i\round{\frac{eB}{2\hbar}}\square{\hg_R\pd_{x}\hg_R^{-1}\pd_{y}\hg_R-\hg_R\pd_{y}\hg_R^{-1}\pd_{x}\hg_R} =i\round{\frac{eB}{2\hbar}}\square{\hg_R\hh_x\hg_R\hh_y\hg_R-\hg_R\hh_y\hg_R\hh_x\hg_R} =i\round{\frac{eB}{2\hbar}} \hat{{g}}^{R(1)} \ ,
\end{align}
where we have suppressed the explicit $\bp$ and $\n$ dependences of the Green functions, and explicit $\bp$ dependencies from the Hamiltonian, and we use $\pd_i$ to represent $\pd_{p_i}$. Here we labeled retarded index $R$ as subscript for the zeroth order GF, i.e. $\hg_R\equiv \hg^R$. We here note that $\pd_{i}\hg_{R}=\hg_{R}\pd_{i}\hh\hg_{R}$ (where $\hh\equiv\hat{\mathcal{H}}/\hbar$), and we have introduced the notation $\hh_i\equiv\pd_i\hh$.

\subsection{Second-order correction}
Then following Onoda {\it{et. al.}}~\cite{onodaPTP06} we compute the second-order correction of the Green function due to $B$. In what follows we label retarded index $R$ as subscript for the zeroth order GF, i.e. $\hg_R\equiv \hg^R$. The second-order correction for the retarded component reads
\beq
\hG^{R(2)}=\round{\frac{e}{2\hbar}}^2F^{ij}F^{i'j'}\hg_R\square{\frac{1}{2}\pd_{ii'}\hg^{-1}_R\pd_{jj'}\hg_R-\pd_{i}\hg_R^{-1}\pd_{j}(\hg_R\pd_{i'}\hg_R^{-1}\pd_{j'}\hg_R)}\ .
\eeq
where we have suppressed the explicit $\bp$ and $\n$ dependences of the Green functions, and explicit $\bp$ dependencies from the Hamiltonian. We use $\pd_i$ to represent $\pd_{p_i}$ and $\pd_{i j}$ to represent $\pd_{p_i} \pd_{p_j}$. 
Defining $\pd_{ii'}\hg_R^{-1}=-\hh_{ii'}$, $\pd_i\hg_R\equiv\hg_{Ri}$ and $\pd_{ii'}\hg_R\equiv\hg_{Rii'}$ and considering electrons in a periodic potential we obtain
\begin{multline}
\label{g2corr}
\hG^{R(2)}=-\round{\frac{eB}{2\hbar}}^2\hg_R\Big[\tfrac{1}{2}\hh_{xx}\hg_R\hh_{yy}+\tfrac{1}{2}\hh_{yy}\hg_R\hh_{xx}-\hh_{xy}\hg_R\hh_{xy}
+\hh_{xx}\hg_R\hh_y\hg_R\hh_y+\hh_{yy}\hg_R\hh_x\hg_R\hh_x-\hh_{xy}\hg_R\hh_x\hg_R\hh_y-\hh_{xy}\hg_R\hh_y\hg_R\hh_x\\
+\hh_x\hg_R\hh_y\hg_R\hh_x\hg_R\hh_y-\hh_x\hg_R\hh_y\hg_R\hh_y\hg_R\hh_x-\hh_y\hg_R\hh_x\hg_R\hh_x\hg_R\hh_y
+\hh_y\hg_R\hh_x\hg_R\hh_y\hg_R\hh_x\Big]\hg_R\equiv \round{\frac{eB}{2\hbar}}^2\hg^{R(2)}\ .
\end{multline}
Hence retarded component of the gauge invariant GF in presence a static homogeneous perpendicular magnetic field (${\bf{B}}=B\hat{z}$) up to $B^2$ contribution can be written as
$\hG^R_\bp(\nu)=\hg^R_\bp(\nu) + i\round{\frac{eB}{2\hbar}} \hat{{g}}^{R(1)}_\bp(\nu) + \round{\frac{eB}{2\hbar}}^2\hg^{R(2)}_\bp(\nu)$.
In the main text we have used $(\w,\bk)$ symbol to represent mechanical frequency and mechanical momentum $(\nu,\bp)$  for convenience.

\section{Orbital magnetic susceptibility at $T=0$}

The grand canonical potential density for the system at zero temperature becomes
\begin{align}
\W&=-\frac{2}S\int_{-\infty}^\infty\frac{d\n}{2\p}\sum_{\bp}\Theta(\mu-\hbar\n){\rm Im}\tr\curly{\round{\hat{\mathcal{H}}_{\bp}-\mu}\hG^R_{\bp}(\n)}
=-\frac{2\hbar}S\int_{-\infty}^\infty\frac{d\n}{2\p}\sum_{\bp}\Theta(\n_F-\n){\rm Im}\curly{\round{\n-\n_F+i\G(\n)}\tr\square{\hG^R_{\bp}(\n)}}\,,
\end{align}
where $S$ is the area of the 2D electron system and $\n_F=\mu/\hbar$. 
Finally, using Eq.~\eqref{g2corr}, the orbital magnetic susceptibility is given by
\beq
\label{chigen}
\chi=\frac{\mu_0e^2}{S\hbar}\int_{-\infty}^\infty\frac{d\n}{2\p}\sum_{\bp} \Theta(\n_F-\n){\rm Im}\curly{\round{\n-\n_F+i\G(\n)}\tr\square{\hg^{R(2)}_{\bp}(\n)}}\,.
\eeq

\subsection{Bloch electrons}

\label{bloch}
Let us start by analyzing the orbital magnetic susceptibility for a generic Bloch electrons with multiple bands. The starting point is Eq.~\eqref{chigen}. In what follows we suppress the explicit $\bp$ and $\n$ dependencies from the GF, and explicit $\bp$ dependencies from the Hamiltonian. In what follows we label zeroth GF as $\hg_R=\hg^R$.  Let us first analyze $\tr\hg^{R(2)}$. Using the fact that $\pd_\nu\hg_R=-\hg_R^2(1+i\G'(\n))$ with $\G'(\n)=\partial_\n\G(\n)$ and using the cyclic property of trace we may write this quantity as
\beq
\tr\hg^{R(2)}=\frac{1}{6}\frac{1}{1+i\G'(\n)}\pd_\nu(B_1-4B_3+4B_4)\,,
\eeq
where
\begin{align}
B_1&=\tr\square{\hh_{xx}\hg_R\hh_{yy}\hg_R-\hh_{xy}\hg_R\hh_{xy}\hg_R}\,,~~~B_2=\tr\square{\hh_{xy}\hg_R\hh_{x}\hg_R\hh_y\hg_R+\hh_{xy}\hg_R\hh_{y}\hg_R\hh_x\hg_R}\,~~~{\rm{and}}~~~ B_3=\tr\square{\hh_x\hg_R\hh_x\hg_R\hh_y\hg_R\hh_y\hg_R}\,
\end{align}
with
\beq
3B_2+4B_3+2B_4=B_1\,.
\eeq
The orbital susceptibility then becomes
\beq
\label{chifinal2}
\chi= -\frac{\mu_0e^2}{12\p S\hbar}\int_{-\infty}^\infty d\n\sum_{\bp} {\rm Im}\curly{F(\n)(B_1-4B_3+4B_4)}\,,
\eeq
where $n_F(\n)=\Theta(\n_F-\n)$ and
\beq
\label{chifinal2ll}
F(\n)= \pd_\n\square{\frac{n_F(\n)(\n-\n_F+i\G(\n))}{1+i\G'(\nu)}}
= n_F(\n)\round{1-\frac{i\G''(\n)(\n-\n_F+i\G(\n))}{(1+i\G'(\n))^2}}-\frac{i\G(\n_F)\de(\n-\n_F)}{1+i\G'(\n_F)}\,
\eeq
with $\G''(\n)=\partial^2_\n~\G(\n)$.

\section{Two-band systems with particle-hole symmetry}

We now apply the formalism from the previous subsection to a two-band system with particle-hole symmetry. For such a system, the Hamiltonian can generically be written as $\hat{\mathcal{H}}_{\bk\s}={\bf{\rm{f}}}_{\bk}\cdot\btau$, where $\btau$ is the vector of Pauli matrices in sub-lattice or band space and ${\bf{\rm{f}}}_{\bk}$ is a three-dimensional vector that depends on the wave-vector $\bk$. If we now introduce the unit vector $\bn_{\bk}\equiv{\bf{\rm{f}}}_{\bk}/\ve_{\bk}$, the Hamiltonian can be re-expressed as $\hat{\mathcal{H}}_{\bk}=\ve_{\bk}\bn_{\bk}\cdot\btau$. The Hamiltonian has two eigenvalues $\ve_{\bk;\al}=\al\ve_{\bk}=\al|\bff_{\bk}|$, where $\al=\pm1$. Suppressing the explicit $\bk$ indexes from ${\bf{\rm{f}}}\equiv (f_x,f_y,f_z)$ and $\ve$ the corresponding eigenvectors are given by
\beq
\label{eigenvecs}
u_{\bk+}=\frac{1}{\sqrt{2\ve}}\cvec{\frac{f_-}{\sqrt{\ve-f_z}}}{\sqrt{\ve-f_z}}\,,\ \ \ u_{\bk-}=\frac{1}{\sqrt{2\ve}}\cvec{-\frac{f_-}{\sqrt{\ve+f_z}}}{\sqrt{\ve+f_z}}\,,
\eeq
with $f_-=f_x-if_y$. Eigenvectors satisfy $\hat{\mathcal{H}}_{\bk} u_{\bk\al}=\al\ve_{\bk} u_{\bk\al}$.
In the diagonal basis, the (free) retarded Green function reads
\beq
g^R_{\bk;\al}(\w)=\left[{\w-\al\ve_{\bk}/\hbar+i\G(\w)}\right]^{-1}\,.
\eeq
If we now introduce the projector operator $\sP_{\al}=(1+\al\bn_{\bk}\cdot\btau)/2$, the retarded Green function in the {\em original basis} can be expressed in terms of the Green function in the diagonal basis through
\beq
\label{projgf}
\hg^R_{\bk}(\w)=\left[{\w-\hat{\mathcal{H}}_{\bk}/\hbar+i\G(\w)}\right]^{-1}=\sum_{\al=\pm}g^R_{\bk;\al}(\w)\sP_{\al}\,.
\eeq
Inter-band coupling arises because {\em derivatives} of the Hamiltonian are not diagonal in the diagonal basis. 

\subsection{Useful identities}

We now derive some useful identities that will help us re-express Eq.~\eqref{chifinal2} in line with~\cite{raouxPRB15}. In the following, the subscript $\bk$ will be kept implicit for brevity. Starting with the Hamiltonian $\hat{\mathcal{H}}=\ve\bn\cdot\btau$, 
its momentum derivatives are given by
\begin{align}
{\hat{\mathcal{H}}}^i&\equiv\pd_{k_i}\hat{\mathcal{H}}=\ve^i\bn\cdot\btau+\ve\bn^i\cdot\btau\,,
~~~~ {\hat{\mathcal{H}}}^{ij}\equiv\pd_{k_i}\pd_{k_j}\hat{\mathcal{H}}=(\ve^{ij}-\ve\bn^i\cdot\bn^j)(\bn\cdot\btau)+\ba^{ij}\cdot\btau\,,
\end{align}
where $\ve^i=\partial_{k_i}\ve$, $\ve^{ij}=\partial_{k_i}\partial_{k_j}\ve$, $\bn^i=\partial_{k_i}\bn$, $\bn^{ij}=\partial_{k_i}\partial_{k_j}\bn$ and $\ba^{ij}=\ve^i\bn^j+\ve^j\bn^i+\ve\bn\times(\bn^{ij}\times\bn)$.

We now look at the effect of the projectors on the Hamiltonian and its derivatives. A very useful formula in this part is
$(\ba\cdot\btau)(\bb\cdot\btau)=\ba\cdot\bb+i(\ba\times\bb)\cdot\btau$
which holds for any vector $\ba$ and $\bb$. We then find
\begin{align}
\sP_\al {\hat{\mathcal{H}}}^i\sP_\al&=\al\ve^i\sP_\al\,,~~~
\sP_\al {\hat{\mathcal{H}}}^i\sP_\bal=\frac\ve2\square{\bn^i+i\al(\bn\times\bn^i)}\cdot\btau\,, \nonumber \\
\sP_\al {\hat{\mathcal{H}}}^{ij}\sP_\al &=\al(\ve^{ij}-\ve\bn^i\cdot\bn^j)\sP_\al\,~~~{\rm{and}}~~~
\sP_\al {\hat{\mathcal{H}}}^{ij}\sP_\bal=\frac12\square{\ba^{ij}+i\al(\bn\times\ba^{ij})}\cdot\btau\,, \label{projectro}
\end{align}
where $\bal=-\al$. 

\subsection{Quantum geometric tensor}

In the following, the subscript $\bk$ will be kept implicit for brevity. 
The gauge-invariant quantum geometric tensor is given by
$Q^\al_{ij}(\bk)=\olap{\pd_i u_\al}{\pd_j u_\al}-\olap{\pd_i u_\al}{u_\al}\olap{u_\al}{\pd_ju_\al}$,
where $\partial_i=\partial_{k_i}$. The Berry curvature is given by the imaginary part of this tensor:
\beq
\W^\al_{ij}(\bk)=i\olap{\pd_i u_\al}{\pd_j u_\al}-i\olap{\pd_j u_\al}{\pd_i u_\al}=-\frac\al2\bn\cdot(\bn^i\times\bn^j)\equiv\al\W_{ij}(\bk) \,.
\eeq
The real part of the quantum geometric tensor, on the other hand, is given by
\beq
M^\al_{ij}(\bk)={\rm Re}\{Q^\al_{ij}\}=\frac14\bn^i\cdot\bn^j\equiv M_{ij}=\mat{M_{xx}}{M_{xy}}{M_{xy}}{M_{yy}}\,.
\eeq

\subsection{Trace involving two Green functions}

In this subsection we consider ($\w,\bk$) as dummy indices for the mechanical frequency-momentum ($\nu,\bp$). 
In the following, $\bk$ and $\w$ dependencies will be kept implicit for brevity. 
Let us now consider $B_1$ from Sec.~\ref{bloch}. Using Eq.~\eqref{projgf}, one can show that 
\beq
B_1=\frac{1}{\hbar^2}\sum_{\al=\pm}g^R_\al g^R_\al S+\frac{1}{\hbar^2}\sum_{\al=\pm}g^R_\al g^R_\bal A\,,
\eeq
where 
$S=(\ve^{xx}-\ve\bn^x\cdot\bn^x)(\ve^{yy}-\ve\bn^y\cdot\bn^y)-(\ve^{xy}-\ve\bn^x\cdot\bn^y)^2$ and
$A=\ba^{xx}\cdot\ba^{yy}-\ba^{xy}\cdot\ba^{xy}={\bf{\rm{f}}}^{xx}\cdot{\bf{\rm{f}}}^{yy}-{\bf{\rm{f}}}^{xy}\cdot{\bf{\rm{f}}}^{xy}-S$, 
with ${\bf{\rm{f}}}^{ij}=\partial_{k_i}\partial_{k_j}{\bf{\rm{f}}}$. If we now define the contravariant geometric tensor, 
\beq
{{M}}^{ij}(\bk) \equiv \mat{M^{xx}}{M^{xy}}{M^{xy}}{M^{yy}} (\bk) =\mat{M_{yy}}{-M_{xy}}{-M_{xy}}{M_{xx}} \,,
\eeq
such that $M_{ij}{{{M}}}^{jk}={\rm det}(M)\de^k_i$, we find
$S=\ve^{xx}\ve^{yy}-(\ve^{xy})^2+4\ve^2\W_{xy}^2-4\ve {{{M}}}^{ij}(\pd_{ij}\ve)$ and
$A=-2\pd_{ij}\round{\ve^2{{{M}}}^{ij}}-4\ve^2\W^2_{xy}+4\ve {{{M}}}^{ij}(\pd_{ij}\ve)$.

\subsection{Trace involving four Green functions} 

In this subsection we also consider ($\w,\bk$) as dummy indices for the mechanical frequency-momentum ($\nu,\bp$). 
And we here label zeroth GF as $\hat{g}_R=\hat{g}^R$.
In the following, $\bk$ and $\w$ dependencies will be kept implicit for brevity. 
The trace involving four Green functions reads
\beq
4B_3-4B_4=4\tr\curly{\hg_R\hh_x\hg_R\hh_x\hg_R\hh_y\hg_R\hh_y-\hg_R\hh_x\hg_R\hh_y\hg_R\hh_x\hg_R\hh_y}\,.
\eeq
Using Eqs.~\eqref{projgf}-\eqref{projectro}, we may re-expressed the above by
\begin{align}
4B_3-4B_4 &=\sum_{\al_1\al_2\al_3\al_4}g_{\al_1}g_{\al_2}g_{\al_3}g_{\al_4}C_{\al_1\al_2\al_3\al_4}
\end{align}
where $g_\al\equiv g^R_\al$ and $\al_{1/2/3/4}=\pm 1$. After some algebra we find
$C_{\al_1\al_2\al_3\al_4}=\frac{\al_1\al_2\al_3\al_4}{\hbar^4}\square{(\bphi^x_1\cdot\bphi^x_2)(\bphi^y_3\cdot\bphi^y_4)-(\bphi^x_1\cdot\bphi^y_2)(\bphi^x_3\cdot\bphi^y_4)}$,
with $\bphi^{x,y}_l=\al_l{\bf{\rm{f}}}^{x,y}+i\bn\times{\bf{\rm{f}}}^{x,y}$ 
and ${\bf{\rm{f}}}^{i}=\partial_{k_i}{\bf{\rm{f}}}$.
Following lengthy algebra further, we arrive at a relatively simple expression,
\begin{align}
-4B_3+4B_4&=\frac{8\ve^2}{\hbar^3}\sum_\al g^2_\al g_\bal(\pd_{ij}\ve_\al)M^{ij}-\frac{4}{\hbar^2}\sum_\al g_\al g_\bal\pd_{ij}(\ve^2M^{ij})+\frac{8\ve^2}{\hbar^4}\sum_\al g^2_\al g^2_\bal[3(\pd_i\ve_\al)(\pd_j\ve_\bal)M^{ij}-4\ve^2\W^2_{xy}]\,,
\end{align}
where $\varepsilon_\al=\al\varepsilon$ and $\bar{\al}=-\al$.
Adding all the terms, we finally arrive at
\begin{multline}
\label{chifinal2gg}
B_1-4B_3+4B_4=\frac1{\hbar^2}\sum_\al g_\al^2\square{\ve_{xx}\ve_{yy}-\ve_{xy}^2-12\ve^2\W^2_{xy}-12\ve_i\ve_j{{{M}}}^{ij}}
+\frac1{\hbar^2}\sum_\al\al\frac{\hbar g_\al}{\ve}\square{12\ve^2\W^2_{xy}+12\ve_i\ve_j {{{M}}}^{ij}-6\pd_{ij}(\ve^2{{{M}}}^{ij})}\,
\end{multline}
where $\ve_i=\ve^i$ and $\ve_{ij}=\ve^{ij}$, and repeated subscripts and superscripts $i(j)$ stand for tensor sum.

\subsection{Orbital magnetic susceptibility of a two-band system at $T=0$}

Using Eqs.~\eqref{chifinal2} and \eqref{chifinal2gg} we obtain orbital magnetic susceptibility of a two-band system at $T=0$ as $\chi=\chi_{{LP}} +\chi_\W +\chi_{{M}}$ with
\Beq
&& \hspace{-4mm} \chi_{LP} = -\frac{\mu_0e^2}{12\p S\hbar^3}   \hspace{-1mm}  \int_{-\infty}^\infty  \hspace{-1mm} d\n \hspace{-1mm} \sum_{\bp\al} 
{\rm Im}\left[F(\n) g_{\al}^2\right] \left(\ve_{xx}\ve_{yy}-\ve_{xy}^2 \right),~~
  \chi_\Omega = -\frac{\mu_0e^2}{\p S\hbar^3} \hspace{-1mm} \int_{-\infty}^\infty \hspace{-1mm} d\n \hspace{-1mm}  \sum_{\bp\al} 
{\rm Im}\left[F(\n) \left( \al\frac{\hbar g_{\al}}{\ve}-g_{\al}^2\right)\right] \left(\ve^2\Omega_{xy}^2\right), \nonumber  \\
&& \hspace{-4mm}    \chi_{{M}} = -\frac{\mu_0e^2}{\p S\hbar^3} \hspace{-1mm} \int_{-\infty}^\infty \hspace{-1mm} d\n \hspace{-1mm} \sum_{\bp\al} 
{\rm Im}\left[F(\n) \left( \al\frac{\hbar g_{\al}}{\ve}-g_{\al}^2\right)\right] \left(\ve_{i}\ve_{j} {M}^{ij} \right)
 +\frac{\mu_0e^2}{2\p S\hbar^3} \hspace{-1mm} \int_{-\infty}^\infty \hspace{-1mm} d\w \hspace{-1mm} \sum_{\bp\al} 
{\rm Im}\left[F(\n) \left( \al\frac{\hbar g_{\al}}{\ve}\right)\right] \left\{\partial_{ij} \left( \ve^2{M}^{ij}  \right) \right\}, \label{tensor}
\Eeq
where summation is over mechanical momentum $\bp\in$ first B.Z. and $\al=\pm 1$, and integration is over mechanical energy $\n$. The function $F(\nu)$ is defined in Eq.~\eqref{chifinal2ll}. The symbols appear in Eq.~\eqref{tensor} we clarify here as $\ve\equiv\ve_{\bp}$, $g_\al\equiv g^R_{\bp;\al}$, $\ve_i\equiv \partial_{p_i}\ve_{\bp}$, $\ve_{ij}\equiv \partial_{p_i}\partial_{p_j}\ve_{\bp}$, $\W_{xy}=\W_{\bp;xy}$ and ${{M}}^{ij}\equiv {{M}}^{ij} ({\bp})$.

In the main text we have denoted mechanical frequency-momentum $(\nu,\bp)$ through $(\w,\bk)$ for convenience.

\subsection{Generalized graphene systems and orbital magnetic susceptibility of it in presence of a wide bandwidth metallic bath  }
\begin{figure}[h]
\includegraphics[width=0.9\linewidth]{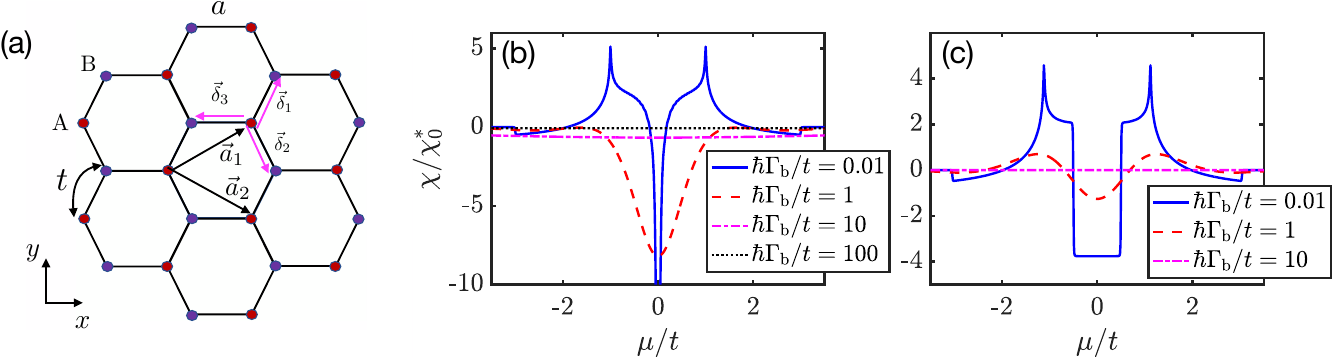}
\renewcommand{\figurename}{Supplementary Figure}
\caption{\label{supple0}{ (a) Schematic of a Honeycomb lattice. $\chi$ vs $\mu$ of (b) a gapless $\De_g=0$ and (c) a gapped $\De_g/t=0.5$ graphene systems attaching to the wide bandwidth metallic bath for various $\G_{\rm{b}}$.  }}
\end{figure} 
The honeycomb lattice of a generalized (gapless and gapped) graphene is schematically represented in Supplementary Fig.~\ref{supple0}(a). The system has two-types of sub-lattice A (red) and B (blue). The nearest neighbor A-B distance is $a$ and hopping amplitude between the nearest neighbor atoms is $t$. The system has two lattice vectors $\vec{a}_{1/2}=\frac{3a}{2}\hat{x}\pm \frac{\sqrt{3}a}{2}\hat{y}$. Then a generalized graphene Hamiltonian then can be expressed as follows
\beq
\hat{H}= \sum_{\bR\in B} \De_g~ B^\dag(\bR)B(\bR) - \sum_{\bR\in A} \De_g~ A^\dag(\bR)A(\bR) +t\left[\sum_{\bR\in A} \sum_{i=1,2,3}A^\dag(\bR)B(\bR  +\vec{\de}_i) +\rm{h.c.} \right],
\eeq    
where $A/B$ represents fermionic annihilation operators, $2\De_g\geq 0$ is the generalized band gap of the system,  h.c. is the Hermitian conjugate, $\vec{\de}_3 = -a\hat{x}$ and $\vec{\de}_{1/2} = \vec{a}_{1/2} +\vec{\de}_3$.
We can now employ the Fourier transformation 
$ A(\bR) =\frac{1}{\sqrt{N}} \sum_{\bk\in {\rm{B.Z.1}}} A(\bk)r^{i\bk\cdot\bR}$,
 and $B(\bR) =\frac{1}{\sqrt{N}} \sum_{\bk\in {\rm{B.Z.1}}} B(\bk)r^{i\bk\cdot\bR}$,
where B.Z.1 stands for the first Brillouin zone. 
The generalized graphene Hamiltonian hence can be expressed as $\hat{H}=\sum_{\bk\in{\rm{B.Z.1}}}\Psi_\bk^\dag\hat{\mathcal{H}}_{\bk}\Psi_\bk,$,
where $\Psi^\dag_\bk= \left[B^\dag(\bk)~A^\dag(\bk)\right]$ and $\hat{\mathcal{H}}_{\bk\s}={\bf{\rm{f}}}_{\bk}\cdot\btau$ where ${\bf{\rm{f}}_\bk}=[f^x_\bk,f^y_\bk,f^z_\bk]$ with $f^{x}_\bk=t\cos(k_xa)+2t\cos(k_xa/2)\cos(\sqrt{3}k_ya/2)$, $f^{y}_\bk=-t\sin (k_xa)+2t\sin(k_xa/2)\cos(\sqrt{3}k_ya/2)$, and $f^{z}_\bk=\Delta_g$.

We now demonstrate wide bandwidth bath coupling effects on the orbital magnetic susceptibilities of the generalized graphene systems. For this we model the system-bath dissipation as $\G(\w)=\G_{\rm{b}}$. Supplementary Fig.~\ref{supple0} (b) and Supplementary Fig.~\ref{supple0} (c) demonstrate $\chi(\mu)$ of a gapless ($\De_g=0$) and a gapped ($\De_g/t=0.5$) graphene systems, respectively, for various $\G_{\rm{b}}$. We find that very strong system-bath couplings in this case can completely quench orbital magnetic susceptibility ($|\chi|\rightarrow 0$) of a generalized graphene system. This is equivalent to the famous Bohr-Van Leeuwen theorem.


%